\documentclass[fleqn,usenatbib]{mnras}
\pdfoutput=1
\usepackage{newtxtext,newtxmath}

\usepackage[T1]{fontenc}
\usepackage{ae,aecompl}

\usepackage{graphicx}	
\usepackage{amsmath}	
\usepackage{booktabs}
\usepackage{wasysym}
\usepackage{bm}
\usepackage{subfigure}

\newcommand{\theobj}{LRLL~31\ }

\newcommand{\mum}{\umu\mathrm{m}\/}
\newcommand{\msol}{\rm\,M_{\astrosun}} 
\newcommand{\rsol}{\rm\,R_{\astrosun}} 
\newcommand{\lsol}{\rm\,L_{\astrosun}} 

\title[Fulcrum wavelength of YSOs]{The fulcrum wavelength of young stellar objects - the case of LRLL~31}

\author[G. R. Bryan et al. ]{
Geoffrey R. Bryan,$^{1}$\thanks{E-mail: gbryan@swin.edu.au (CAS)} Sarah T. Maddison$^{1}$ and Kurt Liffman$^{1}$
\\
$^{1}$Centre for Astrophysics and Supercomputing, Swinburne University of Technology,  Hawthorn, Victoria 3018, Australia\\
}

\date{Accepted 2019 August 18. Received 2019 August 11; in original form 2019 April 30}

\pubyear{2019}

\begin{document}
\label{firstpage}
\pagerange{\pageref{firstpage}--\pageref{lastpage}}
\maketitle

\begin{abstract}
A small subset of young stellar objects (YSOs) exhibit ``see-saw" temporal variations in their mid-infrared SED; as the flux short-ward of a fulcrum wavelength ($\lambda_{f}$) increases the flux long-wards of this wavelength decreases (and vice-versa)  over timescales of weeks to years.  While previous studies have shown that an opaque, axisymmetric occulter of variable height can cause  this behaviour in the SED of these objects, the conditions under which a single $\lambda_{f}$ occurs have not previously been determined, nor the factors determining its value.
Using radiative transfer modelling, we conduct a parametric study of the exemplar of this class, LRLL~31 to explore this phenomenon, and confirm that the cause of this  flux variation is likely due to the change in height of the optically thick inner rim of the accretion disc at the dust sublimation radius, or some other phenomenon which results in a similar appearance.  We also determine that a  fulcrum wavelength only occurs for high inclinations, where the line of sight intersects the accretion disc.  
Accepting that  the disc of LRLL~31 is highly inclined, the inner rim radius, radial and vertical density profiles  are  independently varied to gauge what effect this had on $\lambda_{f}$ and its position relative to the silicate feature  near $10 \micron$. While $\lambda_{f}$ is a function of each of these parameters, it is found to be  most strongly dependent on  the vertical density exponent $\beta$.   All other factors being held constant, only for flatter discs ($\beta < 1.2$) did we find a $\lambda_{f}$ beyond the silicate feature.
\end{abstract}




\section{Introduction}
High resolution imaging in the optical and submillimetre provide detailed views of the structure of discs \citep{bro2015a, ans016a, ave2018a}, neither of these techniques can currently image the inner few au of nearby {young stellar object (YSO)} discs where rocky planets are expected to form \citep{dul010a}. 
The spectral energy distribution (SED) of  YSOs have been found to vary in the mid-infrared (MIR), suggesting variation in the distribution of gas and dust in the inner planet-forming region of the disc \citep{muz2009a, fla011, esp010a, esp012a}. 

Infrared space observations have found that a significant proportion of YSOs vary in the MIR on all measured timescales \citep{cod2014a,reb2015a, fla2016a}.
Between 5-30\% of the YSOs are classified as {\it dippers}, which vary on  timescales of the order of days \citep{bou1999a, mor2011a}. They exhibit deep, colourless dips in brightness of occasionally more than a magnitude, and typically with periods similar to the rotation periods of their star.
 One popular hypothesis  to explain this dipping behaviour is that it is caused by the occultation of the central star by a  segment of the inner wall of the disc warped out of the plane of rotation by interaction with a strong magnetic field \citep{lai2008a}. However, recent studies have cast doubt on whether this can be the full explanation \citep{ans016b}.
 \cite{esp010a} uncovered another  class of MIR variable YSOs which exhibit  aperiodic  ``see-saw" flux variations.
The SEDs of these YSOs have a fulcrum wavelength ($\lambda_{f}$\/), about which   {as  the wavelength weighted flux} $\lambda F_{\lambda}$\/ increases shortward of $\lambda_{f}$\/, it simultaneously decreased for wavelengths longward of $\lambda_{f}$\/ and vice-versa.  {This type of MIR variability is the focus of this paper.}

A variety of mechanisms  have been proposed  by \citet{fla011}  to explain  these see-saw like fluctuations in flux. These include:
an increased accretion rate resulting in an increase in the scale height of the disc's inner wall \citep{dul2001a}; an increase in disc surface density leading to an increased  scale height \citep{dal1998a}; or a stellar wind ablating dust from the disc surface and creating an opaque cloud of dust some scale heights high \citep{fla2010a, kon010a}. Magnetic fields might warp the inner disc out of the rotation plane \citep{lai2008a}  or a small unobserved companion orbiting out of the plane of the disc might drag dust with it \citep{lar1997a}. 
One common factor in all of these scenarios is  the presence of an optically thick occulter blocking the observer's line-of-sight to the star. 

Possibly the simplest model is one where there is an inflation and then deflation of an optically thick axisymmetric inner rim located at the dust sublimation radius \citep{dul010a,don2015a}.  
{The presence of a puffed, opaque rim, or something like it, casting a shadow on the disc external to itself was  originally proposed by \citet{nat2001a}   and expanded upon by \citet{dul2001a} to account for the properties of the SEDs of Herbig Ae/Be stars. 

When the opaque inner rim inflates, it casts a shadow on the disc radially {exterior} to it. 
The region in the shadow will then cool rapidly to the ambient temperature of the stellar environment (typically to tens of degrees Kelvin), provided we ignore the effects of thermal diffusion from the regions still illuminated. 
At the same time, for all but a face-on presentation of the system, the apparent height of the illuminated inner rim increases and the amount of shortwave flux will increase linearly as a function of the rim height. Thus, the result of the disc puffing is to cause a pivot in the SED with an increase in the shortwave flux and a decrease in the longwave flux emitted by the object.
While this model does explain the temporal see-saw variation in the SEDs of these YSOs, it does not explain the presence of a unique $\lambda_{f}$\/  for  these systems.

This paper   seeks  to determine what additional factors must be taken into account  to yield  a  unique $\lambda_{f}$\/ for systems exhibiting this  temporal see-saw behaviour in the MIR and which of these parameters most strongly determines its value. We consider the exemplar of the class of see-saw variables, LRLL~31, to study this phenomenon.

\section{Methods}
We use the Monte Carlo radiative transfer code Hochunk3D \citep{whi2003a, whi2003b, whi2013a}  to model the SED of {cTTs} {to explore key parameters that could cause the see-saw variability in the MIR flux}. Using a parametric disc model, the code calculates the disc temperature and then the SED.
We specify the stellar properties of LRLL~31 (mass, radius and effective temperature), plus the geometry and material properties of the accretion disc.
We use a cylindrical-polar mesh with a logarithmic distribution of cell sizes in the radial direction. 
A Planck blackbody curve (with limb darkening) is used to model the stellar source. 
In our simulations we specify a disc accretion rate, $\dot{M}_{\rm disc}$, instead of a fixed viscosity parameter, $\alpha$, as an accretion rate  can be directly  derived from observations of the Bracket and Paschen absorption lines \citep{muz2009a}, and  allow the code to calculate the disc accretion luminosity, which is  added to the total calculated luminosity of the disc.

Hochunk3D models the accretion disc as two  distinct, inter-penetrating  discs which can have different geometric and material properties.  In our simulations, we model a thin, settled disc, which contains larger dust particles and a more extended disc, which contains smaller particles.   
Both discs are modelled as a mixture of dust and gas, with 20\% of the total disc mass comprising the settled disc of larger grains (following the model of \cite{woo002a}). The remaining 80\% of the total disc mass is found in the extended disc with fine grain parameters derived by \cite{kim094a} from  properties of ISM dust. 

The disc density, $\rho$, is parameterised in Hochunk3D as:
\begin{equation}
 \rho \left(R, z \right) = \rho_{0} \left( 1 - \sqrt{R_{\star}\over{R} }\right) \left( {R_{\star}\over R } \right)^{\alpha} \exp \left\{- {1\over 2} \left[z\over {h\left(R \right)} \right]^{2}\right\}.
 \label{eqn:parametric-density}
\end{equation}
Where  the scale height, $h$, is given by equation \ref{eqn:whitney-scale-height}   
\begin{equation}
h(R) = h_{0} \left( R\over R_{\star}\right)^{\beta},
\label{eqn:whitney-scale-height}
\end{equation}
and where $R$ and $z$ are the radial and vertical cylindrical components, $R_\star$ the stellar radius, $\rho_0$ is the local mid-plane density,  and $\alpha$ and $\beta$  respectively the radial and vertical density profile exponents.

{Following the arguments of \citet{lyn1974a} and \citet{pri1981a}, a characteristic length scale in the $z$-direction, the isothermal scale height, $H_{\rm iso}$ can be derived
\begin{equation}
	H_{\rm iso}(R) = 
	\sqrt{\frac{k_{B}T_{\rm eff} R^3}{\bar{m}\mu G M_{\star}} }
 =\left(
 \sqrt{{k_{B}T_{\rm eff}}\over{\bar{m} {\mu}}}\middle/\sqrt{{GM_{\star}\over{R_{\star}}}}\right)\left( {R \over R_{\star}}\right)^{3\over 2}R_{\star} \, ,
    \label{eqn:isothermal-scale-height}
\end{equation}
wherein $k_B$ is the Boltzmann constant, $\bar{m}$ the mean molecular mass of the gas, $\mu$ the atomic mass unit, $G$ the gravitational constant, and $T_{\rm eff}$ and $M_\star$ the stellar effective temperature and mass. }

{At the stellar photosphere radius the isothermal scale height is:
\begin{equation}
h_{0} =  H_\mathrm{iso}(R_{\star}) =
\left(
 \sqrt{{k_{B}T_{\rm eff}}\over{\bar{m} {\mu}}}\middle/\sqrt{{GM_{\star}\over{R_{\star}}}}\right)     
 R_{\star} 
\end{equation}
and it is this value  that we use  to set the value of the constant $h_{0}$ in equation \ref{eqn:whitney-scale-height}.}

Figure~\ref{fig:trho-in} illustrates the geometry of the inner region of the LRLL~31 system. The primary geometric properties are the radius of the star $R_{\star}$, the dust sublimation radius $R_{\rm sub}$, the co-rotation radius $R_{\rm co}$, and the disc truncation radius $R_{\rm trunc}$.  In the outer region of the disc (Figure \ref{fig:trho-out}) the relevant  geometric quantities  are: the radius to the edge of the inner disc gap $R_{\rm gap}$, the gap  width $\delta R$, and the radial position of the inner rim of the outer disc $R_{\rm od}$. The inclination of the disc to the observer is denoted by  ${i}$. 

\begin{figure*}%
\label{fig:sys_geometry}
\subfigure[]{%
\label{fig:trho-in}%
\includegraphics[width=0.93\columnwidth]{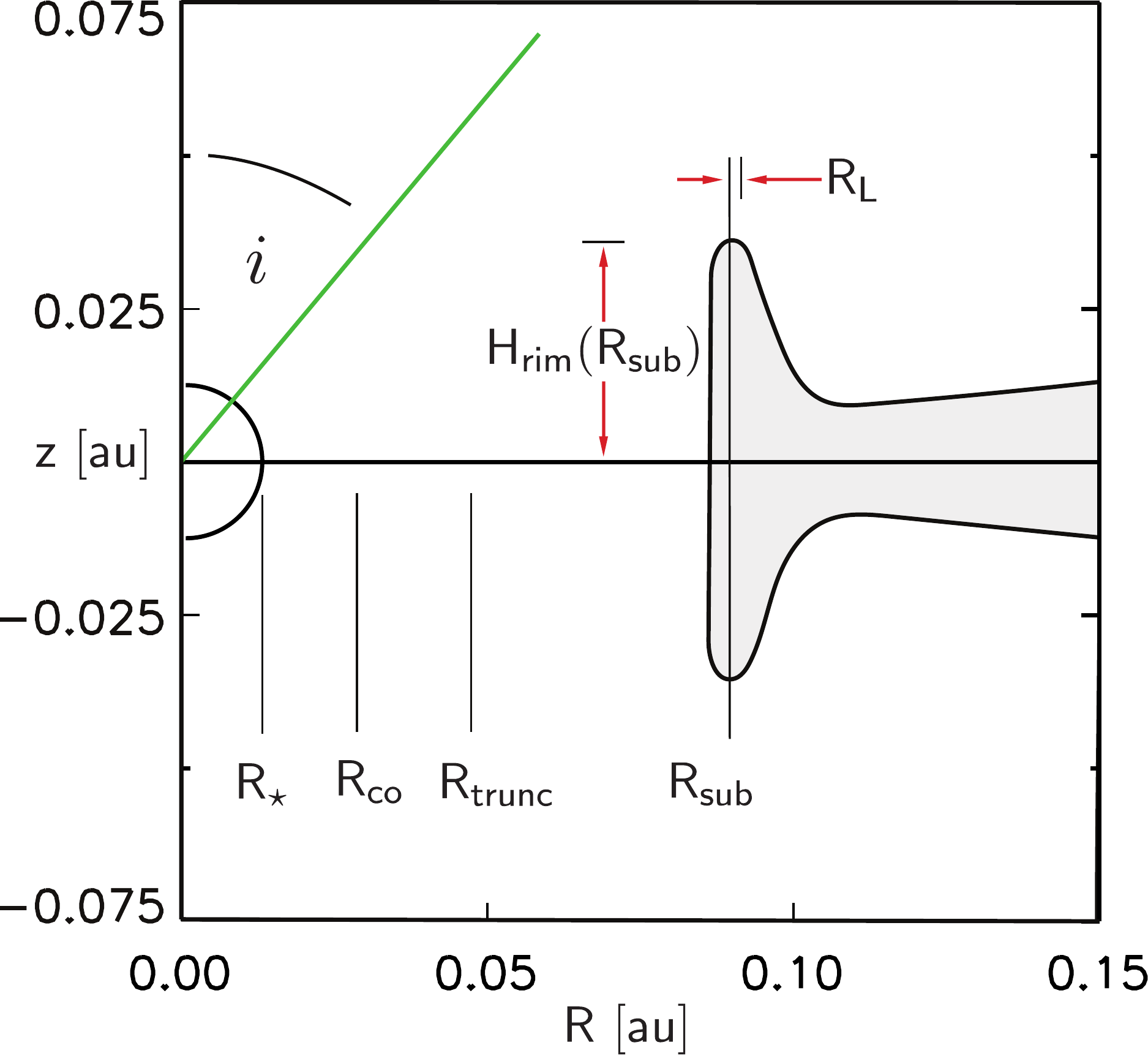}}%
\qquad
\subfigure[]{%
\label{fig:trho-out}%
\includegraphics[width=0.9\columnwidth]{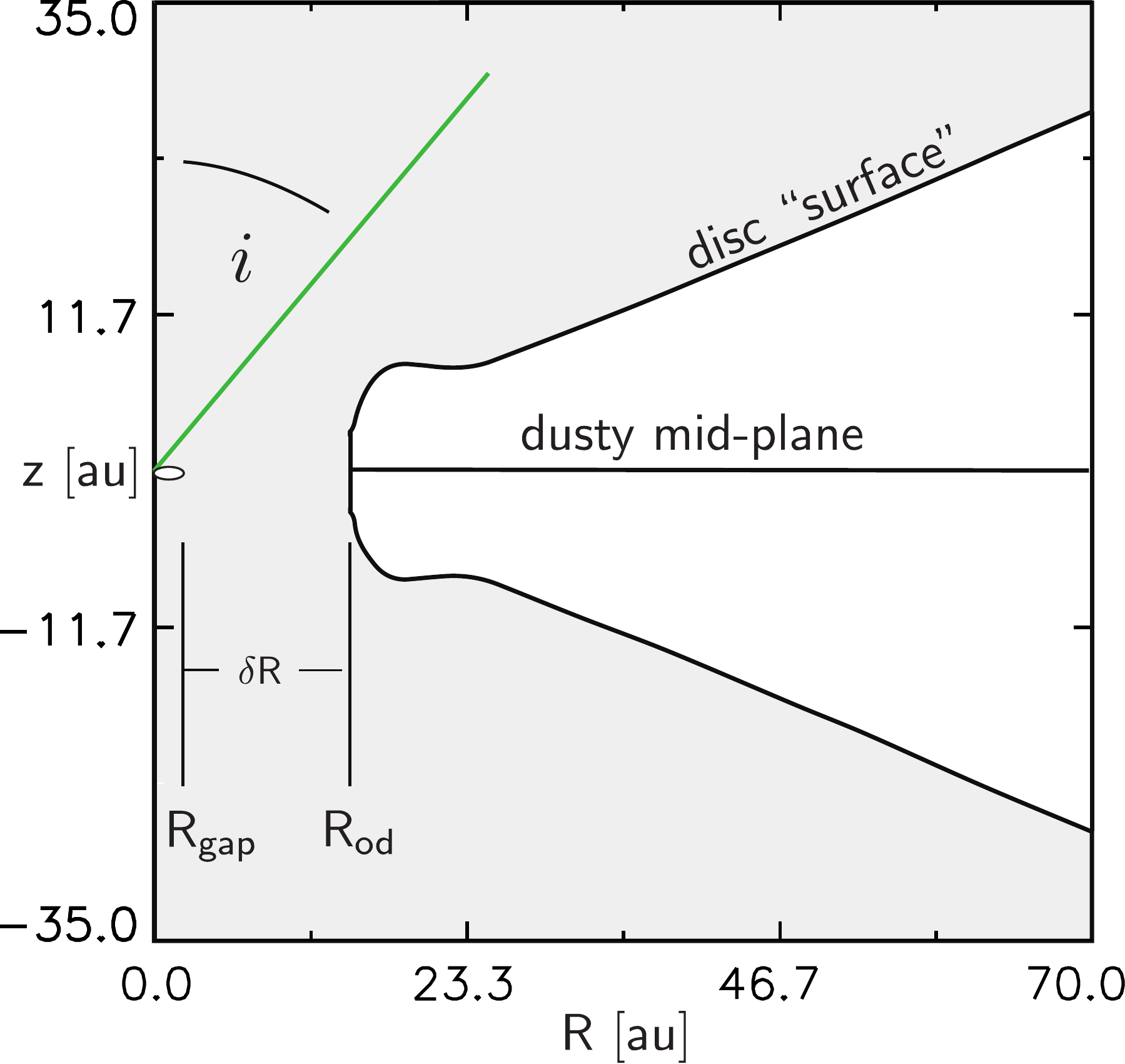}}%
\caption{Schematics of the geometry of LRLL~31 disc.
\subref{fig:trho-in} Inner disc showing the puffed inner rim ($R_{\sf co}$ and $R_{\sf trunc}$ not to scale); and 
\subref{fig:trho-out} outer disc {with a gap of width ${\delta R}$.} 
}
\end{figure*}

The dust sublimation radius $R_{\rm sub}$ is set by the dust sublimation temperature $T_{\rm sub}$ for silicates ($\sim$1,500 K), and material interior to $R_{\rm sub}$ is assumed to be optically thin. Just beyond $R_{\rm sub}$ dust will condense, forming the inner wall of the accretion disc.  Heating of the inner wall by stellar irradiation is thought to puff the inner rim, $H_{\rm rim}$, by some factor greater than unity of the isothermal scale height and also potentially increase the inner rim length in the radial direction, $R_{\rm L}$.

Using a slight modification of the Stefan-Boltzmann relation  due to \citet{whi2004a}, the dust sublimation radius of LRLL~31 can be estimated:
\begin{equation}
R_{\rm sub} = \left( {T_\mathrm{eff}\over{T_{\rm sub}}}\right)^{2.1} R_{\star} \label{eqn:stef-boltz}.
\end{equation}
For $T_{\rm sub}\approx 1,500$\ K, $R_\mathrm{sub} = 16.5 R_{\star}$ and the orbital period at that distance is $P =21.3$d.  This sets a natural time scale for the inner regions of the system: of the order of days to weeks.

Parameters which could vary on timescales of days and weeks and result in fluctuations in the MIR SED on similar timescales include the disc accretion rate, $\dot{M}_{\rm disc}${,} and the inner rim height {$H_\mathrm{rim}$.}

\begin{table}
\begin{center}
\caption{Properties  of LRLL~31.}
\label{tab:stellarprop}
\begin{tabular}{llll}
\toprule \toprule
Parameter & Description & Value & References \\
\midrule
SpType &  Spectral type& G6V  & [1]     \\
$M_{\star}$\/ & Mass   & $1.6 \msol$ &[2], [3]\\
$T_\mathrm{eff}$ & Temperature&   5,700 \rm  K & [2], [6] \\
 $R_{\star}$\/& Radius& $2.3~\rsol$ & [2]\\
 $L_{\star}$\/ & Luminosity&  $5~\lsol$ &[2] \\ 
 $P_{\star}$ & Rotational period  & $3.4$\/ {\rm days} &[2] \\
$ B_{\star}$& Magnetic field strength & 0.15 \rm T & [4] \\
$d$  &  Distance & $315\ {\rm pc}$\/ & [5, [7]\\
$A_\mathrm{V}$ & Extinction & $\sim 8.3$ & [1] \\
\bottomrule
\end{tabular}
\end{center}
[1] \cite{muz2009a}, [2] \cite{fla011}, [3] \cite{esp012a}, [4] \cite{lif2019a}, [5] \cite{pin2014a}, [6] \cite{ken1995a}, [7] \cite{luh2003a}\\
\end{table}

\begin{table}
\begin{center}
\caption{Characteristics of the \theobj accretion disc.  }
\label{tab:accretionprop}
\begin{tabular}{llp{1.8cm}l}
\toprule
\toprule
Parameter & Description & Value  & References\\
\midrule
$\dot{M}_{\rm disc}$\/ &Accretion rate & $0.25-3.32 \times 10^{-8} \msol  {\rm yr^{-1}}$\/  &[1], [2]\\
$M_{\rm disc}$  &Disc mass & $< 0.06 \msol$  & [4]\\
$R_{\rm disc}$ & Maximum  disc radius& $300$~au & [4]\\
$R_{\rm trunc}$ & Truncation radius & $5\rsol$  &[5]\\
$\overline{m}$\/& Mean molecular mass  & 2.3 amu &\\
${i}$\/&Inclination& $> 70^{\circ}$ &  [3]\\
\bottomrule
\end{tabular}
\end{center}
[1] \cite{muz2009a}, [2] \cite{fla011}, [3] \cite{fla2010a}, [4] \cite{esp012a}, [5] \cite{lif2019a}
\end{table}

\begin{table}
\begin{center}
\caption{Additional parameters  in the  fiducial model of LRLL~31.} 
\label{tab:fiducialprop}
\begin{tabular}{lp{3.0cm}ll}
\toprule
\toprule
Parameter & Description & Value \\
\midrule
$\alpha$\/ &{\small radial density  exponent}    & 2.25 \\
$\beta$\/ &{\small vertical density exponent}  &1.25\\
${H}_{\rm puff} $\/ &Additive  rim puffing factor & $0.0$\/ \\
${R}_{\rm L}$\/ &Inner rim length &0.01~au \\
$\delta R$\/ &Disc gap width&0.0~au \\
{\tt FDISK} & {\small $M_{\rm settled}/M_{\rm disc}$ }& 0.2  \\
\bottomrule
\end{tabular}
\end{center}
\end{table}

\subsection{Modelling the exemplar - LRLL~31}
The exemplar of these see-saw variable YSOs is LRLL~31, located in the nebula IC348 of the Perseus star-forming region in Perseus. IC348 is approximately 315~pc from the Sun and contains a young stellar cluster approximately 2 million years old.  \theobj is of spectral type G6V.
\theobj has been observed by multiple observers with the {\it Spitzer Space Telescope} \citep{muz2009a, fla011}.  
Radiative transfer modelling by \citet{esp012a}  suggests it is a pre-transition disc with a gap in its disc of width 14~au.
\cite{muz2009a} found significant infrared variability  on time-scales as short as a week, with the SED pivoting about $\lambda_f = 8.5 \micron$.

Unless otherwise explicitly noted,  we use disc properties  derived from  observations reported in  \citet{muz2009a, fla2010a}  and \citet{fla011} as the basis for our fiducial simulation.

Given our primary goal is to test for the key parameters that could result in a fulcrum in the SED of LRLL~31, rather than try to determine the  best model fit for the SED, we have made a number of simplifying assumptions in our fiducial model. 

We assumed that the occulter (an opaque inner rim wall or some other object ) is axisymmetric. 

We assume that the central star is a blackbody with $T_\mathrm{eff}=5,700$~K and spectral type G6V. While the code does allow for more accurate stellar  atmosphere models, our aim here is to test a relatively simple geometric model rather than make absolutely accurate predictions of the SED of LRLL~31.

Unpublished  observations in the $I$ band by N. Baliber et al.    \citep{fla011}  show small  variations  in the flux which  are hypothesised to be due to the rotation of cool spots  across the stellar surface, which suggest that the rotational period of the star is approximately 3.4 days.
The co-rotation radius, $R_\mathrm{co}$, is the  radius at which the orbital period matches the rotation period of the star at its equator, which} for LRLL~31 it is at approximately 0.03 au.

Magnetic field strengths around cTTs have been deduced to be approximately  1 kG \citep{bou2007b}. 
The truncation radius, $R_\mathrm{trunc}${,} demarcates the edge of the magnetospheric cavity  at which position the ram pressure of the disc material exceeds the support provided by the magnetic pressure. At $R_\mathrm{trunc}$   the disc terminates and the disc  material accretes directly onto the surface of the star in accretion columns. For LRLL~31, adopting  $B \approx 1.5\ \mathrm{kG}$ and $\dot{M}_\mathrm{disc} = 1.6\times 10^{-8} \msol \mathrm{yr^{-1}}$, \citet{lif2019a} calculated $R_\mathrm{trunc} \approx 0.13 \mathrm{au}$.

\citet{esp012a} observed millimetre fluxes from LRLL~31 and following \citet{bec1990a} and \citet{bec1991a} were able to calculate an upper bound on the disc mass of {$ M_\mathrm{disc} < 0.06 \msol$}. We, however, adopt a lower  value of $M_{\star} = 0.01 \msol$.  We also adopt  $R_\mathrm{disc}= 300\  \mathrm{au}$ used by  \citet{esp012a} in their radiative transfer simulations.

LRLL~31 is a pre-transition  object with at least one disc gap occurring between 1  and 15~au \citep{esp012a, pin2014a}. However, for the sake of simplicity, our  initial model assumes no gap.   We will see the effect of the gap in our latter model.

Over the two years post December 2007{,} the extinction $A_\mathrm{V}$ of {LRLL~31} was observed by \citet{fla011} to vary between 8.1 to 9.2  with an average  at $A_\mathrm{V} \sim 8.3$, while the polarisation  was observed to  vary between 7.70\% and 8.44\%. The values of  {the polarization} are high compared to other nearby cTTS and indicate that some portion of the {stellar}  light  passed through the bulk of the   disc{,} suggesting that it is highly inclined to the observer. Following this reasoning  \citet{fla2010a} adopted a value of $i=85^{\circ}$. Previously,  \citet{muz2009a} had also adopted a near edge on orientation for the object.  For this study, we  assume only that the  disc inclination $i > 70^{\circ}$, but for each Hochunk3D simulation run we sweep through 90 values of  the inclination, equi-spaced in $\cos(i)$ between $6.0^{\circ}$ and $89.7^{\circ}$ so as to observe the full effect of varying disc inclination.

Accretion rates for \theobj were deduced from observations of the Pa$\beta$ and the Br$\gamma$ lines by \citet{fla011}. The line luminosities had been shown to be directly correlated with the accretion luminosity \citep{muz1998a}  via:
\begin{equation}
L_\mathrm{acc} = {{ 3 G \dot{M}_\mathrm{disc} M_{\star}}\over{5 R_{\star}} }.
\end{equation}
Over 9 epochs of observation between October 2008 and November 2009, $\dot{M}_\mathrm{disc}$ varied between 0.25 and $3.32 \times 10^{-8} \msol \mathrm{yr^{-1}}$. As the fiducial value of the accretion rate we chose $1.60 \times 10^{-8} \msol \mathrm{yr^{-1}}$ (which occurred on the $8^\mathrm{th}$ November 2008) as a value between the ends of this range.

\citet{fla011} subtracted  the spectrum of the Type G6, WTTS HD  283572 augmented by a Kurucz  model ($T_\mathrm{eff}= 5,750$, $\log g =2.5$) and reddened to $A_{V} = 8.2$  to determine the infrared excess due to the supposed puffed inner rim of LRLL 31. Working on the assumption that the inner rim is optically thick (and so can be accurately approximated by a single temperature  blackbody), they fitted the infrared excess to determine the temperature of the inner rim and the covering fraction of the star. Based on observations over 11 nights between 2005 and 2009 they deduced that $1,540 < T_\mathrm{rim} < 1,940$\ K, with an average of $T_\mathrm{rim} = 1,830$\ K. This is considerably higher than the dust sublimation temperature  ($T_\mathrm{sub} = 1,500$\ K) discussed in \citet{dul010a} and is towards the upper end of the temperature range at which refractory dust species sublime. They derived a  small, dust sublimation radius of 0.05 au for the highest sublimation temperature and  a larger  radius of 0.3 au for the  lowest dust sublimation temperature.  They argued for the feasibility of the higher temperature provided the dust at that radius consists of larger grains (a few $\micron$) which can radiate more efficiently at this wavelength range than smaller grains ($0.1 \micron$). Hochunk~3D does not use $T_\mathrm{rim} = T_\mathrm{sub}$ directly as a simulation input, but rather the $R_\mathrm{sub}$ derived from this temperature. We adopt $T_\mathrm{sub} =1,500$\ K as our fiducial value.

If the measured infrared excess is due to stellar flux reprocessed by an opaque inner puffed rim then fluctuations in this flux will be proportional to the covering fraction of the stellar source by the inner disc wall, which in turn is proportional to the puffed inner rim height $H_\mathrm{rim}(R)$.  The flux due to the wall is $F_{\lambda, \mathrm{rim}}$, where
\begin{equation}
\label{eqn:wall-flux}
F_{\lambda, \mathrm{rim}} \approx {{4\pi R_\mathrm{rim} H_\mathrm{rim}(R_\mathrm{rim}) B_{\lambda}\left( T_\mathrm{rim}\right)} \over{d^{2}}} \sin (i).
\end{equation}
{and $B_{\lambda}(T_\mathrm{rim})$ is the  value of the Planck function at wavelength $\lambda$ at the temperature of the rim $T_\mathrm{rim}$.}

\citet{fla011} has stated that their analysis of the infrared excess shows that $T_\mathrm{rim}$  (and hence $R_\mathrm{rim}$) remained approximately constant during their observations. So a change in the infrared excess by a factor of two  between $\mathrm{31^{st}}$ October and the $\mathrm{4^{th}}$ November 2009 is best explained by an increase in a factor of two in the puffed rim height $H_\mathrm{rim} (R_\mathrm{rim})$. The largest increase in value in $H_\mathrm{rim}(R_\mathrm{rim})$ was inferred to be a factor of 4 between $\mathrm{8^{th}}$ October 2009 and $\mathrm{9^{th}}$ November 2009. In our radiative transfer simulations we simulate fluctuations in the inner rim height  in the range 1 to 7.5 of the unpuffed rim height.

Unless the accretion disc is particularly massive ($M_\mathrm{disc}> 0.1 \msol$) the contribution to the total luminosity due to viscous dissipation is small  (a few percent) and  such passive discs  flare  {so} that the central star is visible from everywhere on the disc's surface. We  adopt the values for the radial exponent  $\alpha$ and the flaring exponent $\beta$  for a  passive, flaring disc  derived by \citet{ken087a}: $\alpha=2.25$ and $\beta=1.25$, for  our fiducial simulation.

\subsection{Simulation suite}

\begin{figure}
\center
\includegraphics[width=0.95\columnwidth]{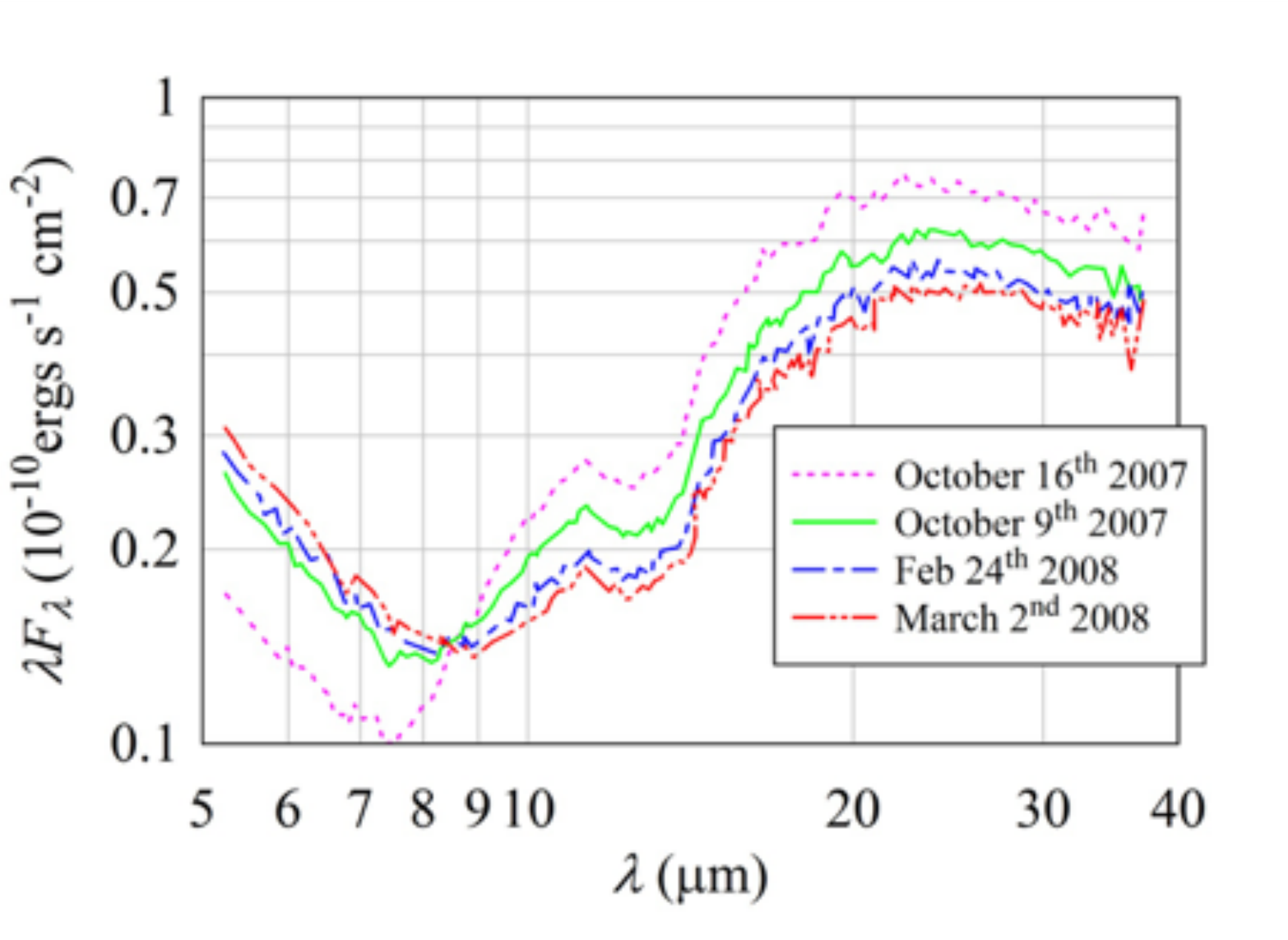}
\caption{The SED of  LRLL~31, with \textit{Spitzer} IRS data from \protect\cite{fla011}.} 
\label{fig:temporalinfraredflux}
\end{figure}

{The stellar and disc parameters of LRLL~31    are given in Tables ~\ref{tab:stellarprop} \&  \ref{tab:accretionprop} respectively, with additional parameters using the fiducial model given in Table \ref{tab:fiducialprop}. In our fiducial model $\dot{M}_\mathrm{disc} =  1.60 \times 10^{-8} \msol \mathrm{yr^{-1}}$ and $M_\mathrm{disc} = 0.01 \msol$.}
 For all our simulations, we compare the resulting simulated SEDs with the \textit{Spitzer} IRS data from \citet{fla2012a} -- see Figure~\ref{fig:temporalinfraredflux}.  We test for the presence or otherwise of a fulcrum point, the value of $\lambda_{f}$, and the magnitude of the weighted flux $\lambda F_{\lambda}$.

\subsubsection{A puffed inner rim}

In our first set of simulations  (Sim. 1; {Table  \ref{tab:simfamilyA}}) we vary the puffed inner rim height scale factor, $H_{\rm puff}$  and the disc inclination to the line of sight, $i$. 

Hochunk3D parameterises the puffed inner rim using the expression (Figure \ref{fig:sys_geometry})
\begin{equation}
H_\mathrm{rim}(R) = h(R) \left[  1 + H_\mathrm{puff}  \exp  \left( - \left({{R-R_{\rm rim}}\over{R_{L}} }\right)^{2}\right)\right]. \label{eqn:puffed-rim}
\end{equation}
Here: $H_\mathrm{rim}(R)$\ is the puffed rim height at $R$,  $h(R)$ is the  scale height of the unpuffed disc at radius $R$, $H_\mathrm{puff}$ is the  puffed inner rim height scale factor {and}  $R_\mathrm{L}$\/ is the characteristic rim length in the radial direction.
$H_{\rm puff}$ varies between 0.0 to 6.5  in 10 equal intervals, and for each of these simulations the inclination varies from being almost face-on ($i=6^{\circ}$) to almost edge-on ($i=87^{\circ}$) on  a grid of 90 inclinations equispaced in $\cos i$\/ (see Table ~\ref{tab:simfamilyA}).

\subsubsection{Varying disc accretion rate, $\dot{M}_\mathrm{disc}$}
Varying the accretion rate in the inner part of the disc is expected to contribute directly to fluctuations in the flux at short wavelengths in the ultraviolet as well as in the infrared.   
Observations of LRLL~31's veiled  Brackett and Paschen lines suggest that it has an irregular accretion rate \citep{muz2009a, fla011}.

{In our second set of simulations (Sim 2; Table \ref{tab:simfamilyA}); we varied} the mass accretion rate between $1.6\times 10^{-8}-1.6\times 10^{-7}$ in 11 uniform steps while keeping all other parameters of the fiducial model constant.  

\begin{table*}
\centering
\caption{Parameters varied in the simulation suite.}
\label{tab:simfamilyA}
\begin{tabular}{llllll}
\toprule\toprule
Simulation & Parameter		& Start value  & End value & No. of increments & Increment\\
\midrule 
 Sim. 1&$H_{\rm puff}$\/& 0.00 &6.50 &  10 & 0.65  \\
 Sim. 2  & $\dot{M}_{\rm disc} (\msol {\rm yr}^{-1})$\/  & $1.6\times10^{-8}$\/ &$1.6\times10^{-7}$\/&  10& $1.44\times 10^{-8}$\/ \\
Sim. 3 & $T_\mathrm{rim}$\/(K) & 1,000 & 2,000& 4 & 250 \\
Sim. 4 &$\delta R\/({\rm au})$\/ & 14.0 &0&  0& 0\\
Sim. 5 &$\alpha$ &2.0 & 2.5&  5 & 0.2 \\
Sim. 6 &$\beta$ & 1.0& 1.3&  6 & 0.05\\
\bottomrule
\end{tabular}
\end{table*}

\subsubsection{ $T_\mathrm{rim}$ and the fulcrum wavelength $\lambda_{f}$}\label{sec:fulcrum_wavelength}

{Assuming that $H_\mathrm{rim}$ and/or $\dot{M}_\mathrm{disc}$ are the parameters whose variation results in a single fulcrum point in the MIR SED,} we next explored the factor (or factors)  which determine the value of ${\lambda_f}$. 

Our principal hypothesis was that  the  stellar temperature is kept fixed, but that the  radius of the inner rim of the accretion disc is no longer at the dust sublimation radius{,} but rather at the much larger radius of the inner disc cavity and that it is the cavity radius which determines $\lambda_{f}$.

As previously noted,  for $R < R_{\rm sub}$ the material of the disc is likely to be optically thin because it will be above the $T_{\rm sub}$.  However, there are many mechanisms  \citep[see][]{koe2013a} that can clear material from the inner zone of the disc: the presence of a small companion \citep{ire2008a},  photo-evaporation \citep{ale2006a, ale2006b}, strong stellar winds \citep{kon2000a} or a recent  star burst  \citep{abr2009a}, eroding the inner edge of the disc  and leaving the outer disc intact, where the temperature is lower than $T_{\rm sub}$\/. 
To explore this  possibility, we ran a suite of simulations where we kept  $T_{\rm eff}=5,700$~K (the fiducial value) while varying the inner rim temperature between 1,000--2,000\ K. We used the relation equation~\ref{eqn:stef-boltz} to calculate  the radius at which $T_{\rm rim}$ took on these temperatures. For each temperature, 11 simulations were run for puffed inner rim scale factors $H_\mathrm{puff}$ between 0.0 and 6.5  (see Table \ref{tab:rimtempvarying}).  {These  simulations mimic the initial set  (Sim. 1) in which the inner rim puffs and deflates due to some as yet undetermined cause, but with a different inner rim radius.}
\begin{table}
\centering
\caption{Additional parameters varied in the inner disc rim temperature simulations (Sim. 3), $T_\mathrm{eff} =5,700$\ K.}
\label{tab:rimtempvarying}
\begin{tabular}{lll}
\toprule \toprule
$T_{\rm rim} [{\rm K}]$\/& $R_{\rm rim} [{\rm au}]$\/  & $H_\mathrm{puff}$ \\ 
\midrule
1,000 &  0.3474 &0 -- 6.5  \\
1,250 &  0.2223&0 -- 6.5  \\
1,500 &  0.1544&0 -- 6.5 \\
1,750 &  0.1134&0 -- 6.5  \\
2,000 &  0.0860 &0 -- 6.5 \\
\bottomrule
\end{tabular}
\end{table} 

\subsubsection{Modelling the effect of disc structure on the SED and the occurrence of $\lambda_{f}$}
To explore more generally the presence or otherwise of a fulcrum wavelength in LRLL~31 and the values it  could take,  we investigated the effect of  changes in other parameters which do not vary on a temporal  scale of  {days or weeks,} and which can be expected to be unique to  different {cTTs.}
These parameters include: the radial  density exponent $\alpha$ and the disc flaring exponent $\beta$ and the presence or absence of a gap in the disc.
 
{We explored the effect of varying the radial density exponent, $\alpha$, on  either side of the fiducial value, $[2.0 < \alpha < 2.45] $}.
{As the discs of cTTs are generally known to flare we selected a range of values for the flaring exponent $\beta$  in our simulations such that the disc  varied between being flat, $\beta=1.0$, and having a slightly more flared profile, $\beta=1.30$, than the fiducial value of $\beta=1.25$. In our simulations $\alpha$ and $\beta$ were varied independently.}

{As our principal aim was to examine  what the effect of the gross structure of the disc had on the existence or otherwise of a fulcrum point and on $\lambda_{f}$, the fiducial simulation included no gap. However, it has been deduced that LRLL~31 is a transition disc with a 14 au gap in its inner disc. Consequently we have also modelled the SED of LRLL~31 where such a gap is present, varying the puffed inner rim height factor, $H_\mathrm{puff}$ in the range: $0 < H_\mathrm{puff} < 6.5 $.}

\section{Results}

\subsection{Varying the puffed inner rim scale factor, $\mathrm{H_{puff}}$}\label{sec:rimheight}

\begin{figure*}
\center
\includegraphics[width= 1.80\columnwidth]{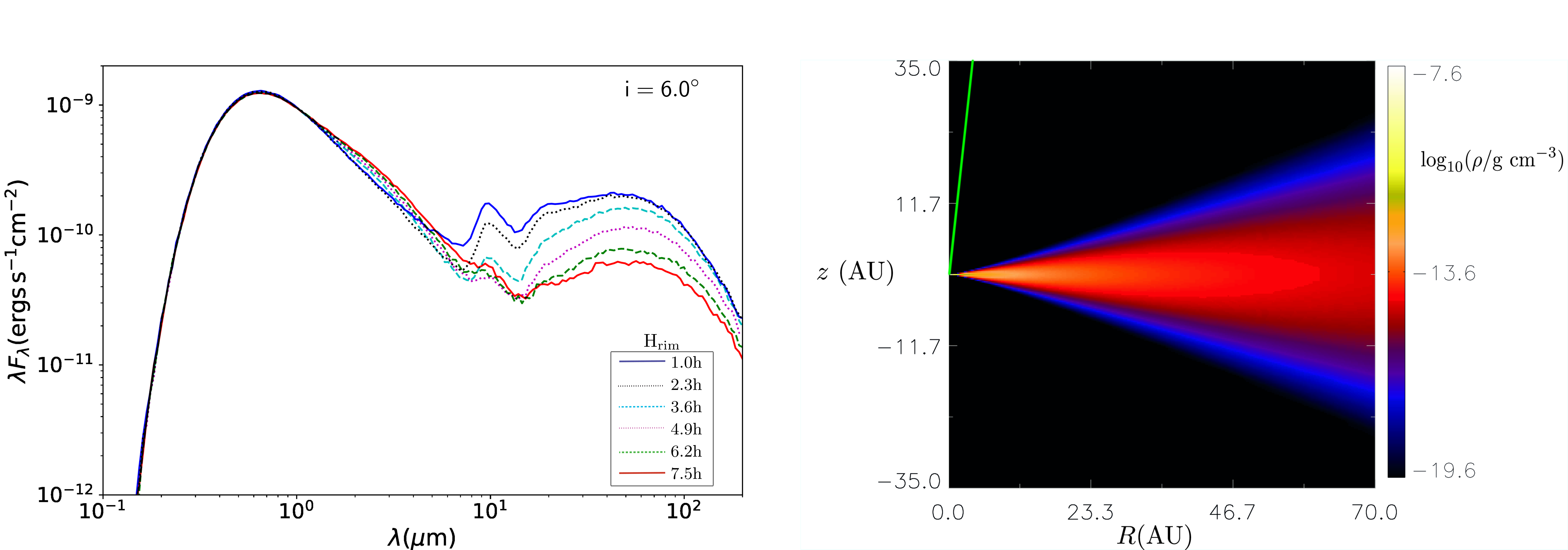}
\caption{Left: {The SEDs} of \theobj\/ for a variety of puffed inner rim heights, $H_{\rm rim}$, for a {near face-on system with inclination ${i} = 6.0^{\circ}$. Right: The modelled disc density with the line of sight superimposed in green} {(data from Sim.1)}}
\label{fig:sim_sed06}
\end{figure*}

We varied the {inner rim height over 11 values (Table \ref{tab:simfamilyA}). Figure~\ref{fig:sim_sed06} shows the resulting SEDs for a near face-on disc with ${i} = 6.0^{\circ}$),  where the puffed inner rim height scale factor varied from $H_{\rm puff} = 0.0 - 6.5$.
Varying $H_\mathrm{rim}$ results in a  see-saw variation in the SED near $8~\micron$\/.  { For increasing $H_\mathrm{rim}$} the flux short-ward of $8 \micron$\/ increases and the flux  long-ward of $8 \micron$\/ decreases.  Note that for this almost face-on presentation, there is no single pivot (or fulcrum) point but rather a series of cross-over points spread between $7$\/ and $15 \micron$\/. The diagram on the right of Figure ~\ref{fig:sim_sed06} shows contour plots of the density of the parametric disc, with the (green) line showing is the line of sight.}

\begin{figure*}
\center
\includegraphics[width= 1.80\columnwidth]{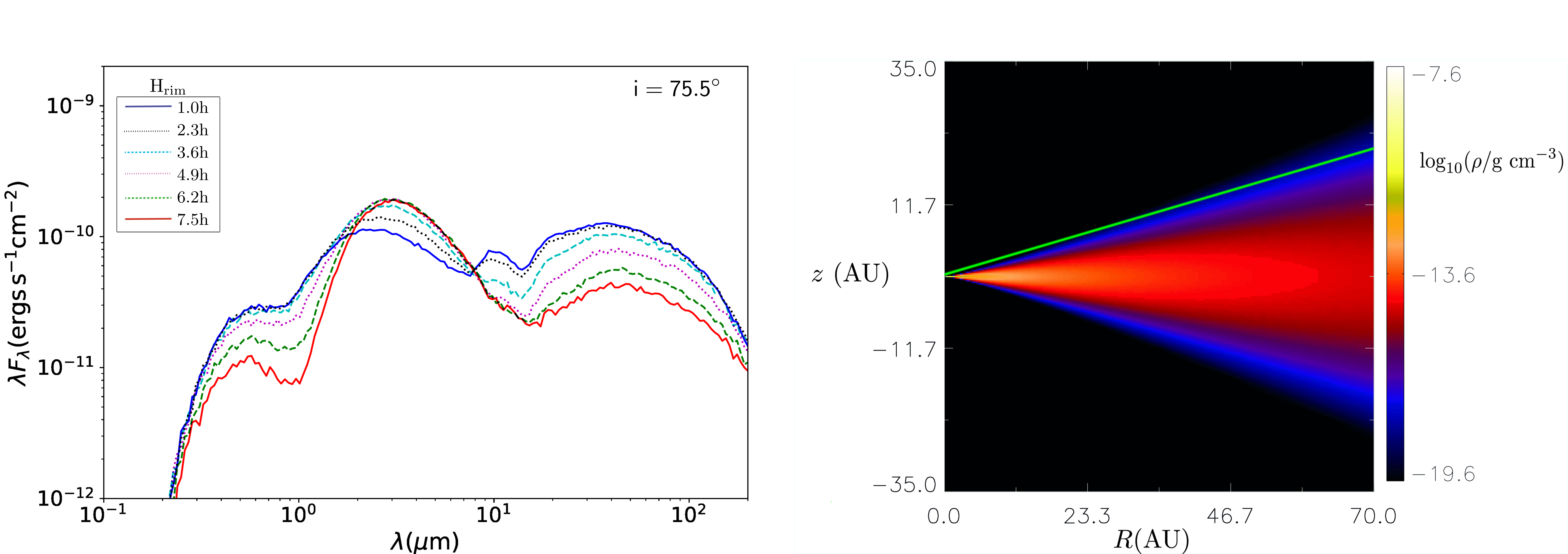}
\caption{Same as Figure~\ref{fig:sim_sed06} for a near edge-on disc with inclination ${i} = 75.5^{\circ}$ (Sim. 1)}
\label{fig:sim_sed67}
\end{figure*}

As we increase the disc inclination, the magnitude of the photospheric peak decreases {while  the peak in the SED due to the inner wall of the  disc  increases in magnitude and becomes prominent.} It is only when  the line of sight begins to graze the {disc surface} at  $\approx 75^{\circ}$ that a single fulcrum wavelength or pivot point appears near $8\micron$\/ {(Figure \ref{fig:sim_sed67})}. As the inclination increases further and the line of sight cuts through the disc,  a pivot point is still observed at near $8 \micron$, though the magnitude of the flux at that point rapidly decreases. For inclinations above $82^{\circ}$, {the simulations become increasingly noisy and it is difficult} to determine the presence (or otherwise) of a pivot point.

\begin{figure}
\center
    \includegraphics[width=0.9\columnwidth]{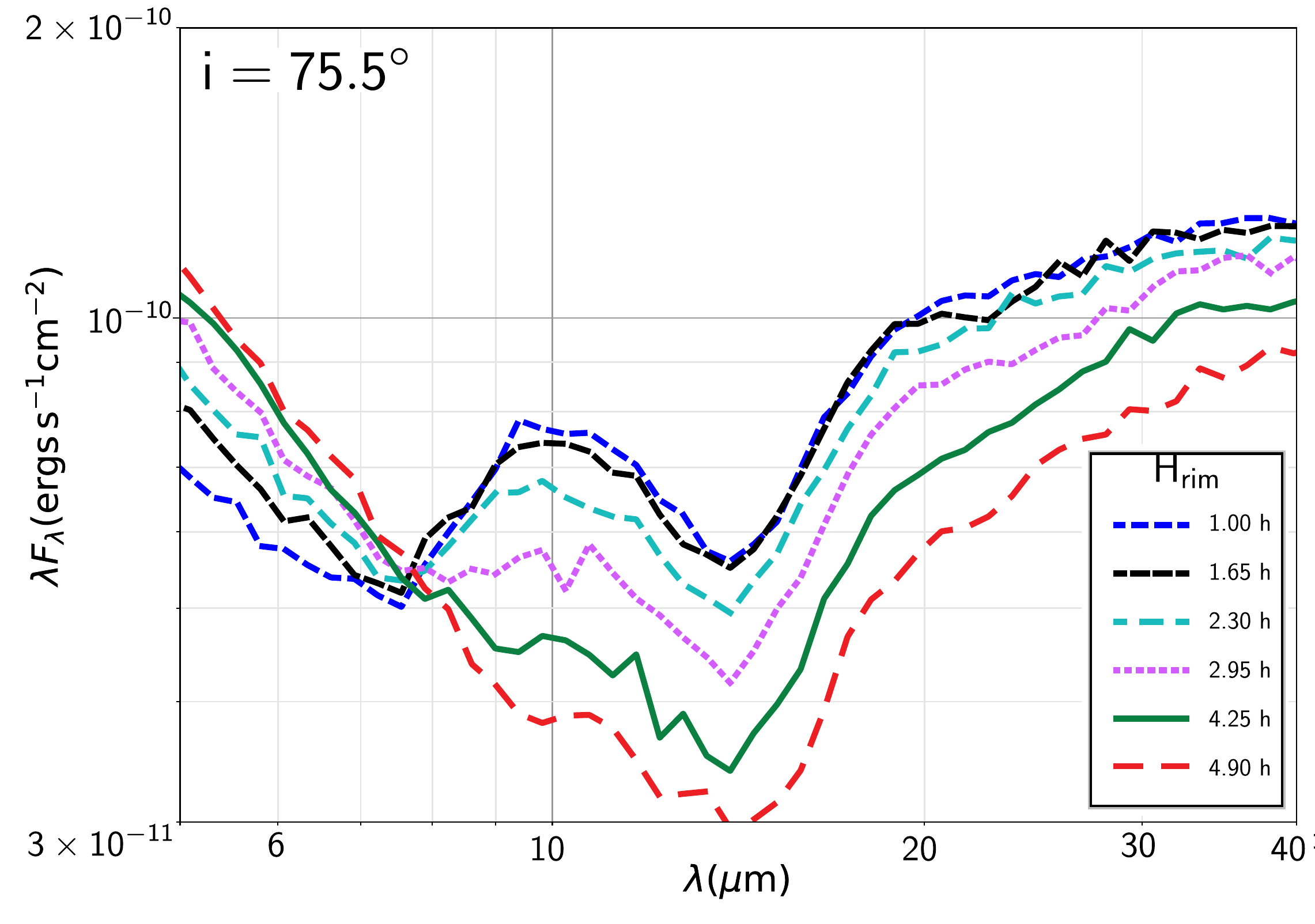}
    \caption{{Mid-infrared portion of the SEDs of LRLL~31, modelled as a full disc viewed near edge-on ($i=75.5^{\circ}$), for a range of $H_{\rm rim}$ values with a fulcrum point near $7.7 \micron$}    (Sim.1) }
    \label{fig:flaherty_comp_figure}
\end{figure}

{Figure~\ref{fig:flaherty_comp_figure} shows the MIR portion of the SED between $5\micron$ and $40\micron$}, {with $i=75^{\circ}$} - no reddening has been applied during post-processing.  The key feature of the plot is the presence of a fulcrum point, $\lambda_{f}$, near $8\micron$ {shortward} of the $10\micron$ silicate feature. This is in general agreement with the {LRLL~31 SED} shown in Figure~\ref{fig:temporalinfraredflux}.

\subsection{Varying the mass accretion rate, \protect{$\dot{\mathrm{M}}_\mathrm{disc}$}}

\begin{figure}
\begin{center}
 \includegraphics[width=0.9\columnwidth]{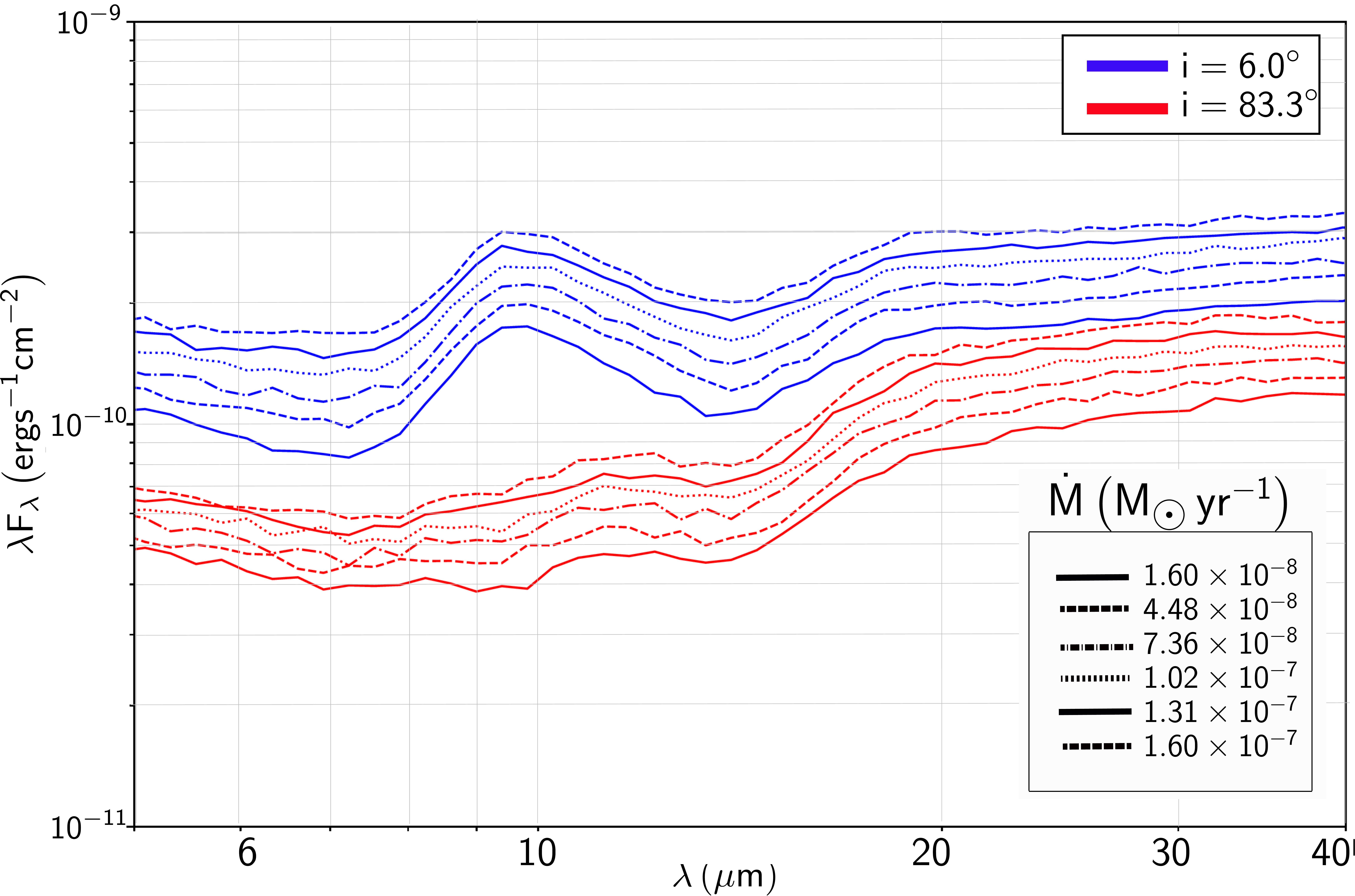}
\caption{MIR SED of LRLL 31  as a function of the mass accretion rate, $\dot{M}$, for an almost face-on (blue)  and edge-on (red) disc {(Sim. 2).}}
\label{fig:sedmdot}
\end{center}
\end{figure}

Figure~\ref{fig:sedmdot} shows the results of simulations {which vary the mass accretion rate between $\dot{M}= 1.6\times10^{-8}$ -- $1.16 \times10^{-7} ~\msol  \mathrm{yr}^{-1}$\/  for two different inclinations: almost face-on with $i=6.0^{\circ}$, and almost edge-on with $i=83.3^{\circ}$  in which the line of sight passes through the disc. In both cases, increasing the accretion rate increases the flux without appreciably changing the shape of the SED.}  Note  the {almost face-on} case has a pronounced $10 \micron$\ silicate feature,  while for the almost edge-on case the silicate feature has disappeared due to the absorption of the bulk of the disc.

\subsection{Fulcrum wavelength as a function of inner rim temperature, $\mathrm{T}_\mathrm{rim}$}
\begin{figure*}
\centering
\includegraphics[height=6.9cm]{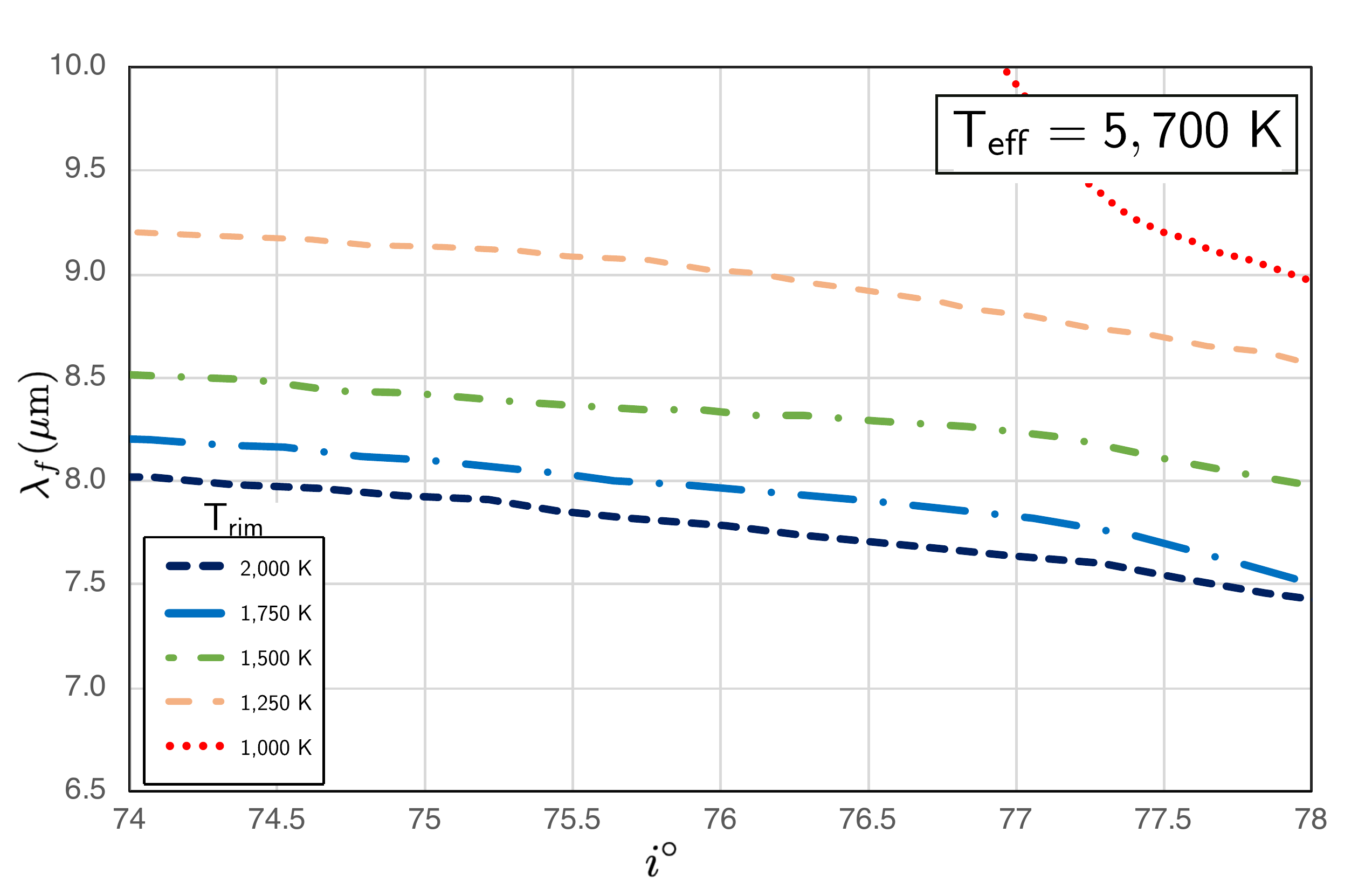}%
\caption{Fulcrum wavelength $\lambda_{f}$ as a function of $T_\mathrm{rim}$ for  $T_\mathrm{eff} =5,700$\ K{, (Sim. 3)} }
\label{fig:lambda_vs_t_rim}
\end{figure*}

Figure \ref{fig:lambda_vs_t_rim}  shows the effect of varying the temperature of the inner puffed rim for a fixed stellar temperature on $\lambda_{f}$ as a function of disc inclination. For a given inclination,  as one  increases the temperature of the inner rim, $\lambda_{f}$.
For inner rim temperatures between 1,250 and 2,000K, $9.2 \micron > \lambda_{f} > 7.4 \micron$.

{It is possible that  some cTTs will have more refractory dust species  located in the inner region of the  disc \citep{dul010a} and that this will lead to  a higher $T_\mathrm{sub}$. It's also possible that as the  disc evolves, the inner disc truncates   and  the disc inner rim will be  located further out  at larger radii and  lower temperatures.}

{If the inner wall rim moves outwards and is cooler,  then the inner wall component  of the SED  will move closer to the silicate feature. If the inner wall rim moves inwards and is warmer,   then inner wall component   will move to shorter $\lambda$ and away from the silicate feature. This should cause the intersection of these two components of the SED - which will be at  approximately  the fulcrum wavelength   - to move towards longer or shorter  wavelengths respectively.   This hypothesis was tested by running simulations for a fixed  $T_\mathrm{eff} = 5,700$\,K, but in which the temperature of the inner wall ($T_\mathrm{rim}$) is varied between $1,000$\,K and $2,000$\,K.  Figure   \ref{fig:lambda_vs_t_rim} depicts  the results of these simulations.}

{We see, that the $\lambda_{f}$ vs. $i$ curves of fixed $T_\mathrm{rim}$ are convex, that they do not intersect, and that for $T_\mathrm{rim} < 1,500$\,K have  larger  values of $\lambda_{f}$ than on the fiducial curve, and for $T_\mathrm{rim} > 1,500$\,K have smaller values of $\lambda_{f}$ than on the fiducial curve. }

\subsection{Gapped disc model}\label{sec:rimheight2}
\begin{figure*}
\begin{center}
 \includegraphics[width=1.80\columnwidth]{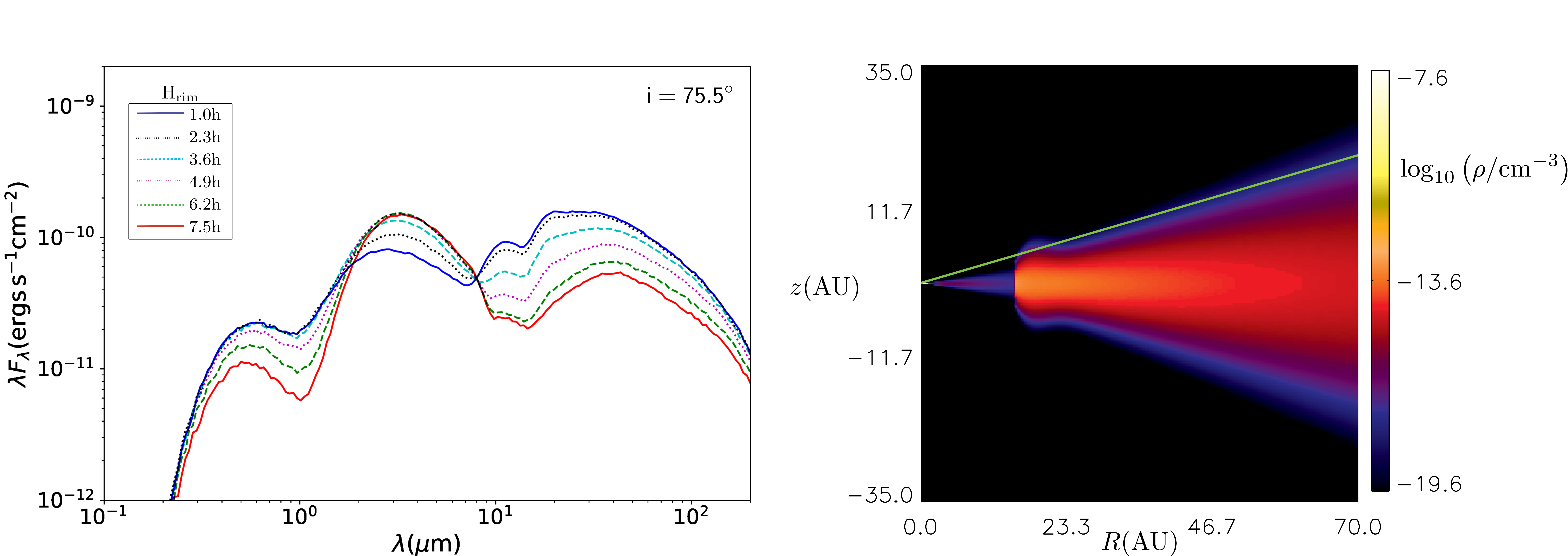}
\caption{SEDs of LRLL 31 {modelled with a gap between 1-15~au with a viewing angle of $i = 75.5^{\circ}$ (Sim. 4)}}\label{fig:sed_gap_out_67}
\end{center}
\end{figure*}

\begin{figure}
\center
\includegraphics[width=0.9\columnwidth]{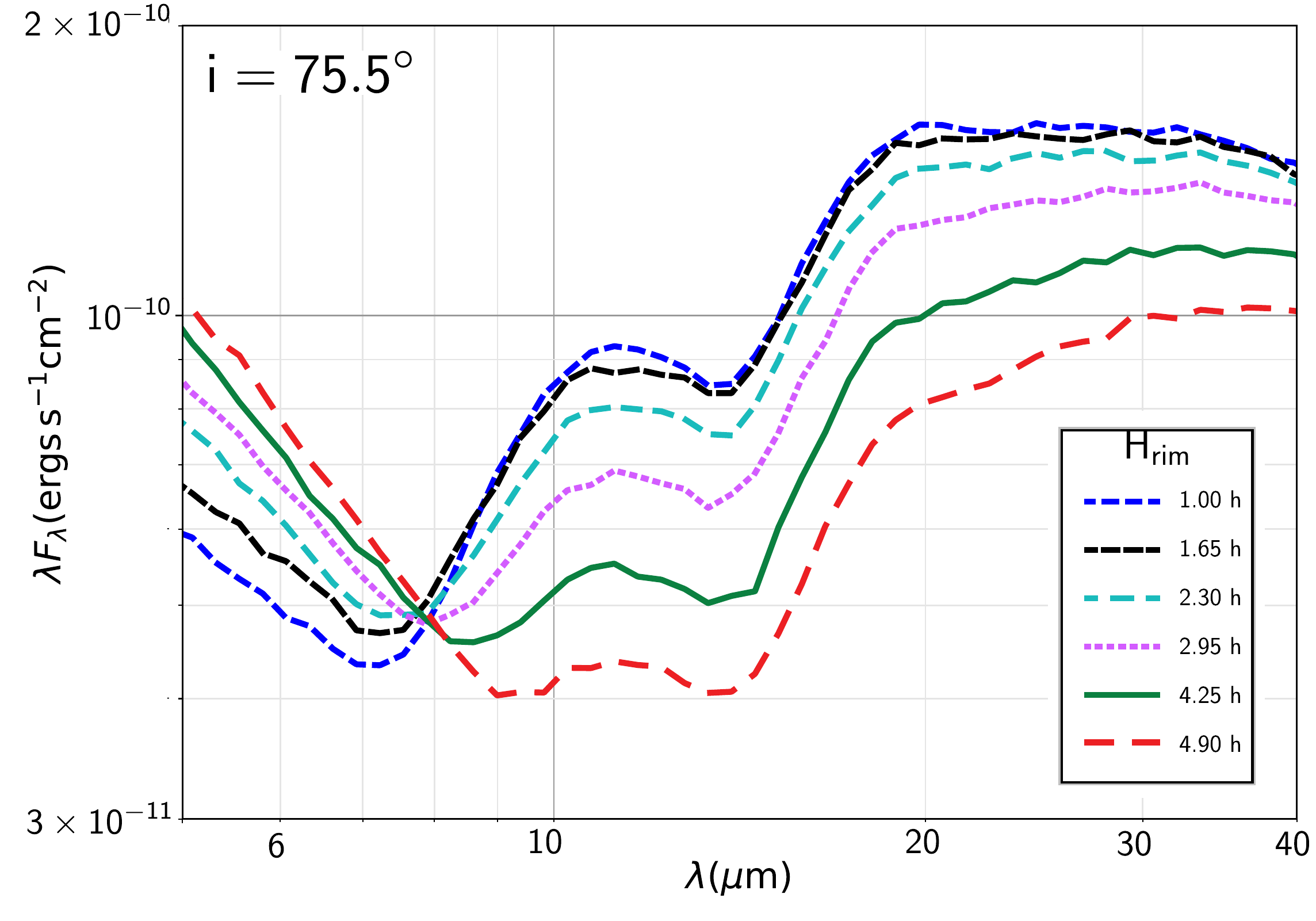}
    \caption{Same as Figure \ref{fig:flaherty_comp_figure} with the disc modelled with a gap between 1-15~au  and displaying a fulcrum point of   $7.9 \micron$ (Sim. 4).}
    \label{fig:flaherty_comp_gap}
\end{figure}

\begin{figure*}%
\centering
\subfigure[]{%
\label{fig:inc_i000}%
\includegraphics[height=5.95cm]{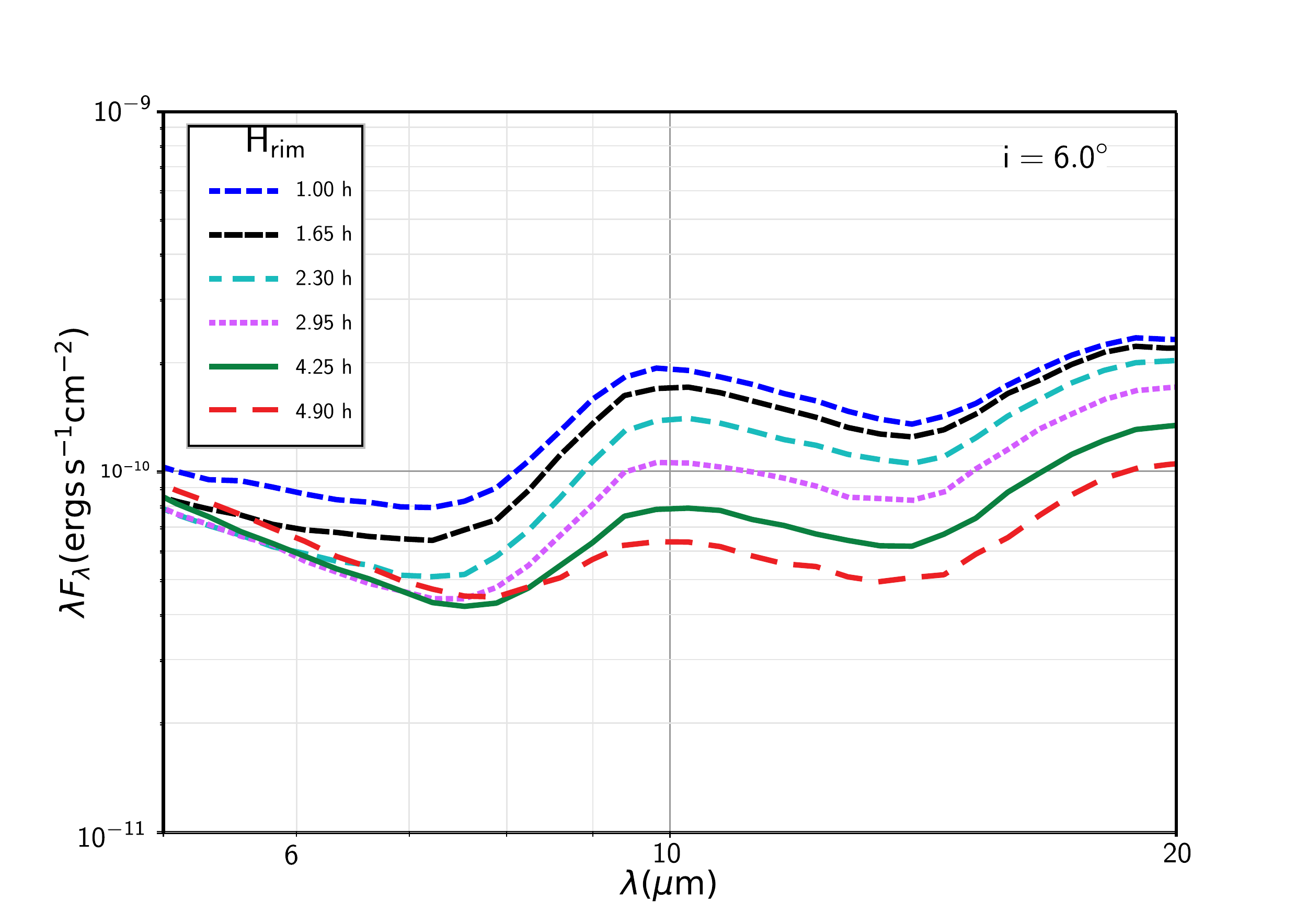}}%
\qquad
\subfigure[]{%
\label{fig:inc_i026}%
\includegraphics[height=5.95cm]{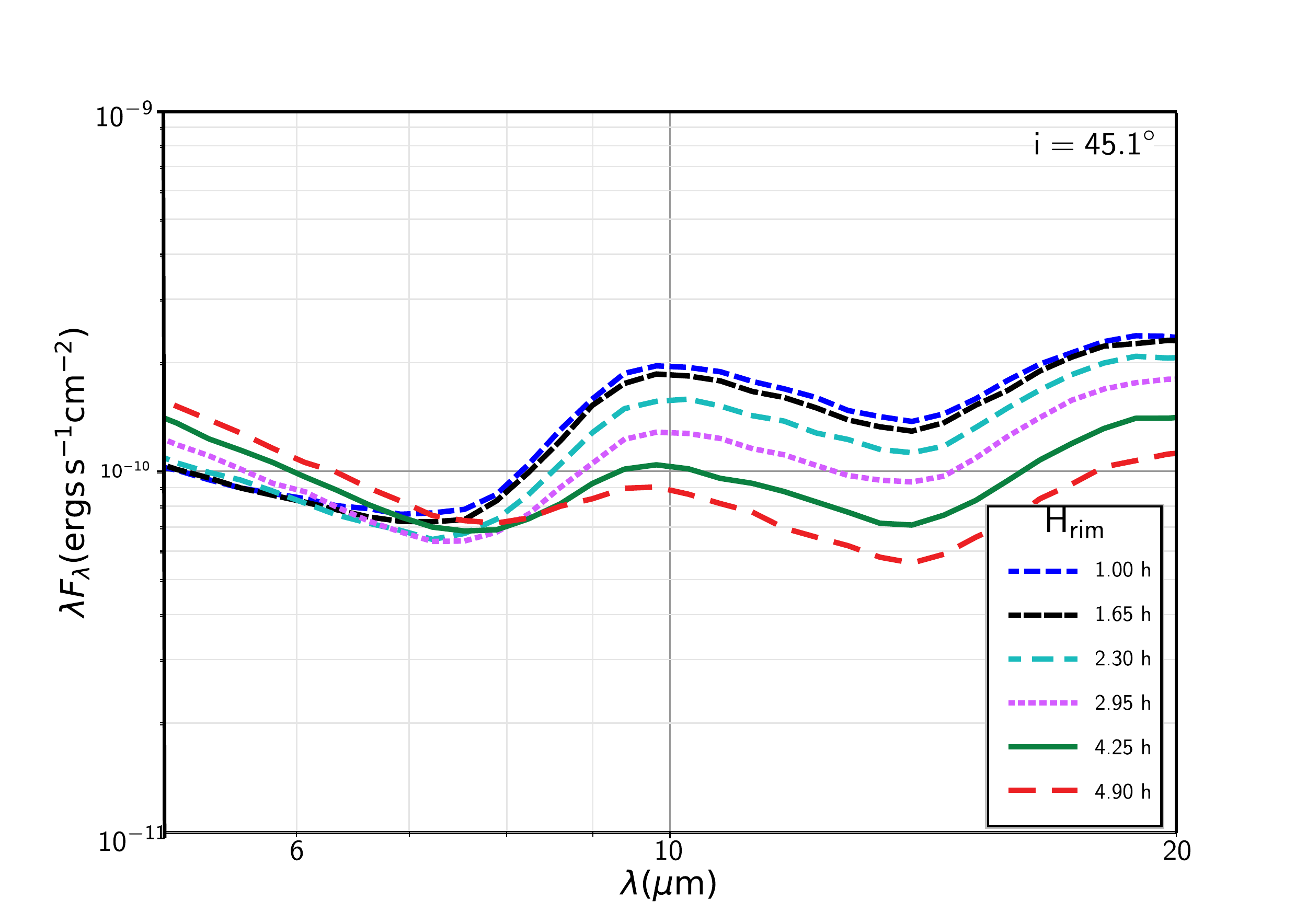}}%
\\
\subfigure[]{%
\label{fig:inc_i045}%
\includegraphics[height=5.95cm]{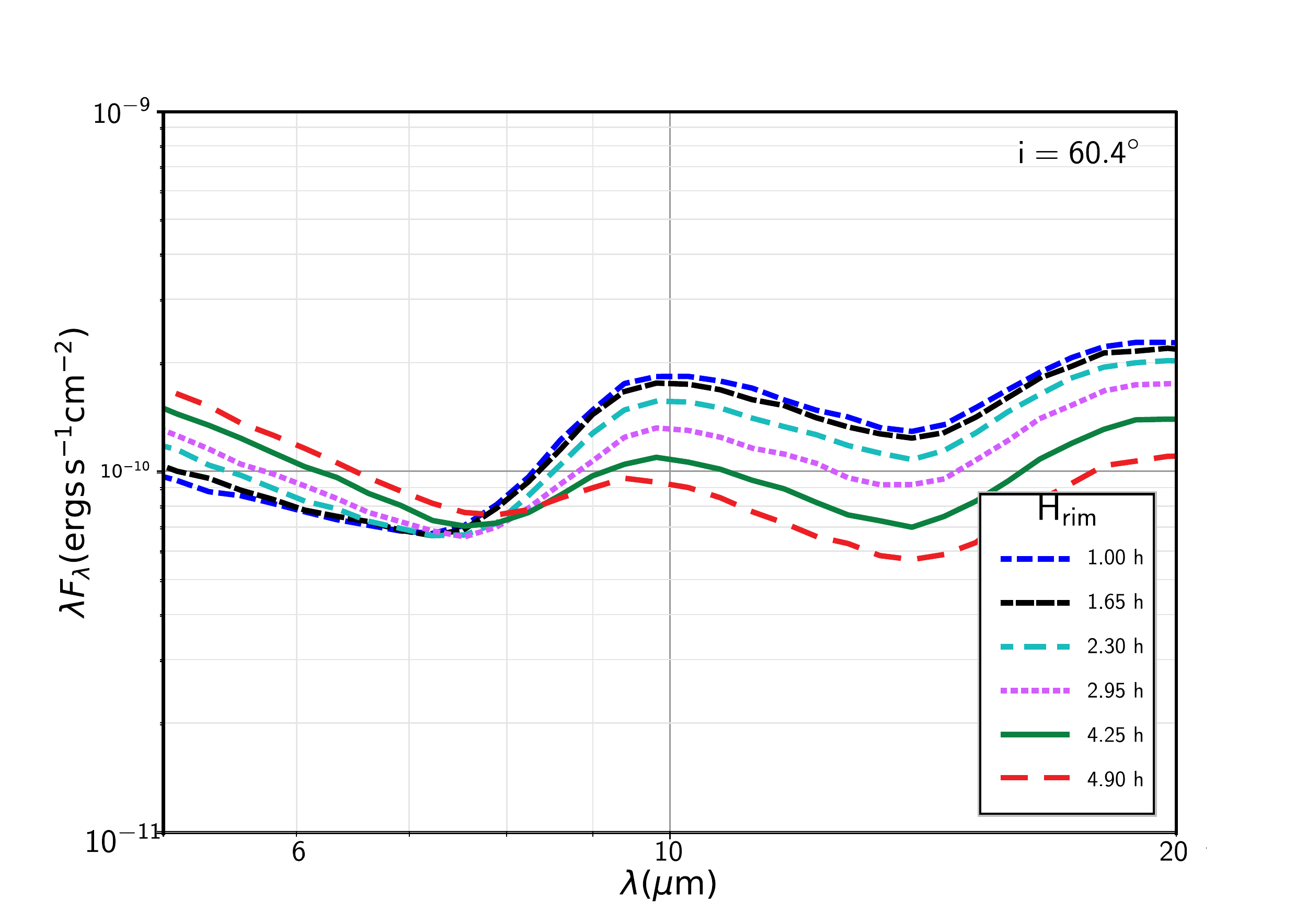}}%
\qquad
\subfigure[]{%
\label{fig:inc_i067}%
\includegraphics[height=5.95cm]{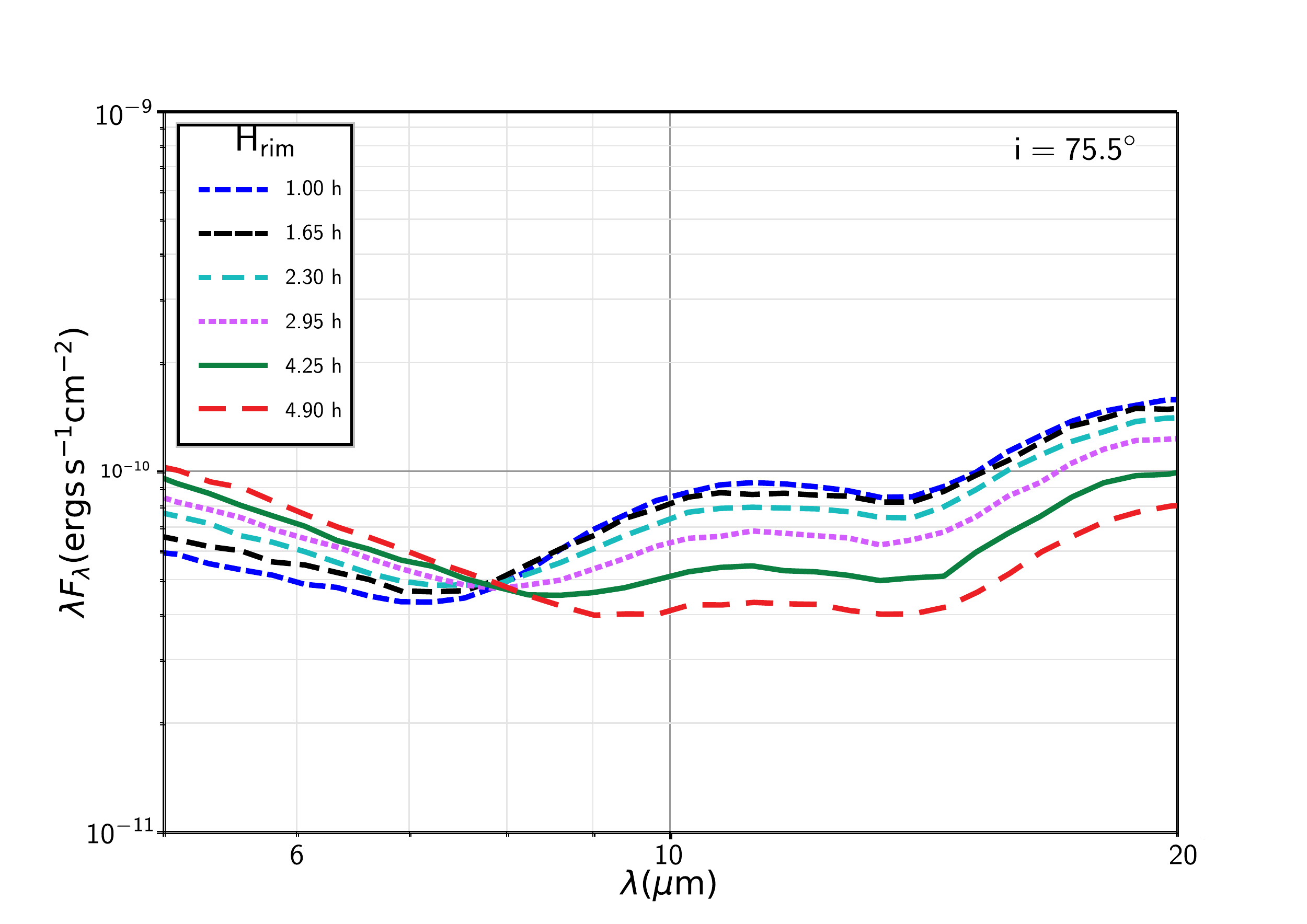}}%
\caption{MIR portion, for a gapped disc (gap  1 - 15 au),  of the SED for four inclinations: 
\subref{fig:inc_i000} $i = 6.0^{\circ}$;
\subref{fig:inc_i026} $i = 45.1^{\circ}$;
\subref{fig:inc_i045} $i = 60.4^{\circ}$; and 
\subref{fig:inc_i067} $i = 75.5^{\circ}$ for each of six puffed rim heights  $(1h < H_\mathrm{rim} <  4.9h )$. Note how the intersections of the curves move long-wards in wavelength as $i$ increases and how the flux does not decrease substantially until the line of sight begins to intersect the disc ($i > 70^{\circ}$).}
\label{fig:gapped_disc}
\end{figure*}

The disc surrounding \theobj has {a gap from approximately 1-15 au \citep{esp012a, pin2014a}.}  Figures~\ref{fig:sed_gap_out_67}  and \ref{fig:flaherty_comp_gap} show the resulting SED for a range of puffed rim-heights, $H_{\rm rim}$ when $i=75.5^{\circ}$. {Again, as displayed in Figures \ref{fig:gapped_disc} \subref{fig:inc_i000}, \subref{fig:inc_i026},  \subref{fig:inc_i045} \& \subref{fig:inc_i067}, we find for $i < 75^{\circ} $ there is see-saw behaviour in the SED as the inner rim height changes, however there is no single fulcrum point until $i \approx 75^{\circ}$ } 

Comparing the MIR portion of the SED for the gapped disc model (Figure \ref{fig:flaherty_comp_gap})  to the same wavelength range of the {full disc model (Figure \ref{fig:flaherty_comp_figure})} and the observational data (Figure \ref{fig:temporalinfraredflux}) we note that the addition of the gap has the effect of: (i)  moving $\lambda_{f}$ from $7.7 \micron$ to $7.9 \micron$, closer to the observed value of $8.5 \micron$; moving the silicate peak in the SED from $\approx 10 \micron$ to $11.2 \micron$ which is close to the observed value of $11 \micron$, and  finally moved the peak in the SED due to the outer wall from $\approx 36 \micron$ to $\approx 20 \micron$, which is closer to the observed position of $\approx 23.8 \micron$.

Comparing the SEDs in Figures. \ref{fig:sim_sed67} and \ref{fig:sed_gap_out_67}, {we note that $\lambda_{f} \approx 8 \micron$ does not change significantly between the full and gapped case, and that the general shape of the SEDs are very similar but that the peak in the curve beyond the silicate feature is slightly more pronounced for the gapped disc.} In the right hand pane of Figure~ \ref{fig:sed_gap_out_67} which displays the density distribution, we see that  exterior to the gap, the disc profile flares to form an outer wall and it is the additional flux from this cooler, outer wall that enhances the flux near  $20\micron$.   {The presence of a 14 au gap in the inner disc improves the match between the simulated SED and that arising from observations.}

\subsection{Varying the disc density profile} 
\begin{figure*}
\begin{center}
 \includegraphics[width=1.80\columnwidth]{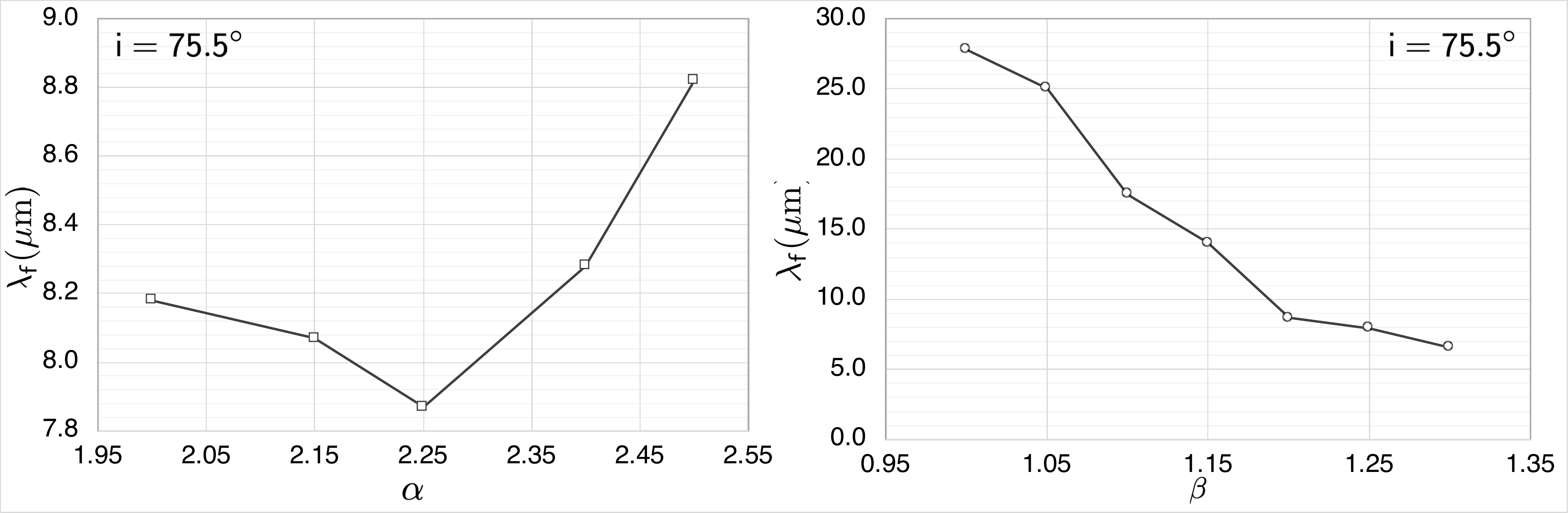}
\caption{Variation in {the fulcrum wavelength, $\lambda_{f}$, as (a): A function of the radial density exponent $\alpha$ {(Sim. 5)}, and (b) the  vertical density exponent $\beta$ {(Sim. 6)} for a disc inclination $i=75.5^{\circ}$.}}
\label{fig:lambdavalpha}
\end{center}
\end{figure*}

{In simulations 5 and 6 we vary the exponents $\alpha$ and $\beta$.}
Figure~\ref{fig:lambdavalpha}(a) shows the fulcrum wavelength as a function of the radial density exponent $\alpha$ for a disc inclination of $75.5^{\circ}$.  The fiducial value of $\alpha = 2.25$ {is varied between 2.0 and 2.5, which results in variations in $\lambda_{f}$  between $7.9 \micron <  \lambda_{f} < 8.8 \micron $.}

{Figure~\ref{fig:lambdavalpha}(b) shows significant variation in $\lambda_{f}$ $(28 \micron >  \lambda_{f} > 7 \micron)$ as the the vertical disc exponent $\beta$ varies  between 1.0 and 1.3 (where the fiducial value is $\beta=1.25$). The value of $\lambda_{f}$ is very sensitive to the degree to which  disc flares,} {with flatter  disc profiles moving $\lambda_{f}$ to longer wavelengths.}

\section{Discussion}

\subsection{Factors determining the presence of a $\lambda{_f}$ }
{The primary goal of this work was  to determine which parameters contribute to the presence of a unique fulcrum wavelength in the SED of this object.}
From  the observations  depicted in Figure \ref{fig:temporalinfraredflux}, we note that the MIR flux varies substantially even within a week  (between $9^\mathrm{th}$ and $16^\mathrm{th}$ October 2007). Therefore,  only  quantities capable of varying substantially on time scales as short as a week can be considered as a potential cause of  the see-saw behaviour.  Of the parameters  varied in simulation, only the rim-height and the disc accretion rate can alter substantially on this timescale.
{Furthermore, the results of our simulations suggest that it is variations in the inner rim height, for highly inclined discs,  that principally contributes to the existence of a unique fulcrum wavelength, not the disc accretion rate.}  {This connection between high disc inclination and the presence of a  fulcrum point has not  previously been identified.}

Other than that it is high, the inclination $i$ of the  disc of LRLL~31  is poorly constrained by observation  to lie between $70 - 85^{\circ}$. So in our first set of simulations we swept over both  rim height and inclination.  For low to moderate  inclinations the  disc does not  obscure the star  and the stellar flux  should remain constant. 

As the rim height is increased, the flux intercepted by the inner disc wall  from the star will increase nearly linearly with its height. If we also assume that the inner rim wall is optically thick,  then, if the wall is at temperature $T_\mathrm{sub}$,  the flux from the  wall will have the form of a blackbody of temperature $T_\mathrm{sub}$.  As the height of the rim increases the radius of the shadow on the disc lengthens and the temperature of the portion of the disc in shadow drops. This reduces the amount of longer wavelength radiation irradiated  by the shadowed disc and causes the longer wavelength portion of the SED to pivot downwards to the right as the rim height increases. 
The $10 \micron$ silicate feature will be in emission.  The form of the SED predicted by these qualitative arguments agrees with the results of the simulations for small inclination ($i = 6^{\circ}$) seen in Figures \ref{fig:sim_sed06} and  \ref{fig:gapped_disc}. Note that for this small angle of inclination there is  no single $\lambda_{f}$ but a set  of crossing points located  between $5$ and $10 \micron$. 

For high inclinations,  as the inner rim height increases the line of sight intersects the puffed inner rim and the stellar photosphere starts to be obscured. The contribution of the photosphere to the total SED quickly falls away, with the emission peak in the SED near $3 \micron$\/ due to the enlarged inner rim becomes much more prominent and the shadow cast on the disc exterior to the inner rim causes less flux to be emitted.  As the rim height increases, the line of sight intercepts a greater density of dust and the $10 \micron$ silicate feature moves from emission almost to absorption - see Figures \ref{fig:sim_sed67} and \ref{fig:gapped_disc}. Near $i=74^\circ$ the line of sight intersects the surface of the disc and it is only at that point that a single $\lambda_{f}$\/ emerges (Figure \ref{fig:gapped_disc}). {For $74^{\circ} < i < 82^{\circ}$ there  is a single $\lambda_{f}$, for $i > 82^{\circ}$ the results of the simulations are noisy and it is difficult to clearly identify $\lambda_{f}$}.

\citet{cra2008a}  and \citet{whi2003b} have  noted that for high inclinations, where the stellar source is observed through the bulk of the disc, the {SED of class II YSOs become difficult to distinguish from that of class I YSOs.}  
{Class I YSOs  typically have a double humped SED with the shorter wavelength peak due to the stellar object viewed at high extinction and a second longer wavelength peak due to the envelope.  We found that our simulated SEDs began to take on this form at $i \approx 82^{\circ}$.} 

{If it is assumed that the inclinations of discs of this class of MIR variables are uniformly distributed between $0^{\circ}$ and $90^{\circ}$  then   we can make a very rough, first  estimate of the fraction of stars in this class that  have a fulcrum point.  This fraction should equal the range of inclinations for which the objects will be observed to have a fulcrum point divided by the range of inclinations for which the SED has the unambiguous form  associated with a class II YSO. Extrapolating from   our simulations of LRLL~31 we find: $f = {{82^{\circ} - 74^{\circ}}\over 82^{\circ}} \approx  0.1$.  That is, about 10\% of cTTs similar to \theobj would appear to an observer to have a fulcrum point. }

\subsection{The link between $\lambda_{f}$ and disc structure}

The second goal of this study was  to determine how the disc structure influences  $\lambda_{f}$  and  the position of the fulcrum point relative to the $10 \micron$ silicate  feature.  A number of the MIR variable cTTs studied in \citet{esp010} have  $\lambda_f > 10 \micron$ and so we explore whether our simulations can shed some light on which physical parameters may cause this shift.  

Variations in the dust sublimation temperature $T_\mathrm{sub}$ or  the  dust sublimation radius $R_\mathrm{sub}$,  cause a small  variation in $\lambda_{f}$ for  high disc inclinations. We show that the smaller $T_\mathrm{sub}$, the larger $R_\mathrm{sub}$ and the larger $\lambda_f$ (see Figure \ref{fig:lambda_vs_t_rim}). 
 However, it is possible that some other physical mechanism could lead to the inner rim of the disc being eroded, leading to a larger $R_\mathrm{rim}$ (such as the presence of a  companion  interior to the dust sublimation radius \citet{lar1997a}).  If this was the case, our simulations suggest that this could lead to $\lambda_f > 10 \micron$. 
 
Indeed, there is  indirect evidence from ALMA and  VLA observations  that planetary companions have carved gaps in the inner disc of  the transition disc  object  GM Aur giving the appearance of an enlarged inner cavity \citep{mac2018a}. 

While SED modelling by \citet{esp012a} and \citet{pin2014a} suggests LRLL~31 hosts a 14 au gap in the inner disc, our simulations show that a gap does not produce a significant change in $\lambda_f$ in our simulations.  On that basis we would not expect gaps  of smaller width to significantly effect the value of $\lambda_f$. This raises the question of whether there is any observational evidence for  YSOs with full discs, i.e.  discs with no gaps, exhibiting this see-saw behaviour. 
\citet{fla2012a}  and \citet{esp012a} identified LRLL 2 as a full disc YSO with a SED pivoting in the MIR.
\citet{esp010} published Spitzer observations of   RY Tau showing that it has a pivoting SED  while \citet{gar2019a} has identified it as a full or near full-disc object. \citet{esp010} also claimed that the SED of the YSO  ISO 52 is  best explained by a full disc.

Our simulations suggest that the flaring parameter $\beta$ is important in determining the fulcrum wavelength $\lambda_{f}$.
Semi-analytic,   accretion disc models such as those developed by \citet{ken087a} and \citet{chi097a} place constraints on the value of the flaring exponent $\beta$ for passive, flaring discs (that is, for those discs in which viscous dissipation is not a significant source of energy, and for which the central star is visible from the surface of the entire disc).
For $\beta < 1.0 $  the outer disc is self-shadowed by the inner disc (i.e. it does not flare), while  $\beta=1$ represents a linear disc. 
\citet{ken087a} derived a minimum value of $\beta=1.125$ for a thin disc and a maximum value, (for a flared disc) of $\beta=1.25$.   \citet{chi097a} derived a maximum value of $\beta = 1{2\over7} \approx 1.29$.   
\citet{dul2001a} demonstrated that the presence of an inner disc cavity and the shadow of a puffed inner rim on the outer disc could explain the presence of a NIR bump in the SED of HAeBes, while in later papers \citep{dul2004a, ise2005a, vanb2005a}, the effects of dust-settling and a puffed, but curved inner rim on the flaring of the outer disc and the shadow cast upon it were further investigated. However, these later researches did not alter the limits on $\beta$ for the outer disc established in the earlier work.

The recent development of new instruments capable of spatially resolving nearby discs in the infrared  has enabled observations which have allowed the direct calculation of $\beta$. Table \ref{tab:betavalues} shows values of $\beta$ derived from observations of accretion discs around a selection of  cTTs and HAe/Bes \citep{ave2018a, lag2006a, dif2009a, kre2016a, rag2012a, wol2017a}. Resolved scattered light imaging and or otherwise observationally constrained SED modelling was used to determine $\beta$. These values range from the substantially self-shadowed (MWC 325, $\beta=0.85$) to the surprisingly flared (UX Ori, $\beta=1.79$). However, it should be noted that values of $\beta$ listed in Table  \ref{tab:betavalues}, cannot be easily compared with each other as the instruments used to make these measurements sample very different wavelength regions and hence different depths  in and radii of their respective discs. 
\begin{table}
\caption{Values of the flaring exponent $\beta$ obtained from  observations of a selection of T Tauri  (cTTs) and Herbig Ae /Be  (HAe/Be) stars}
\label{tab:betavalues}
\begin{center}
\begin{tabular}{llll}
\toprule \toprule
Object & $\beta$  &Type &Ref. \\
\midrule
MWC 325& $0.85  \pm 0.13$ &HAe/Be& [4]   \\
RXJ 1615&  $1.116 \pm 0.095 $ &cTTs& [1]  \\
 HD 102964  & $1.26  \pm 0.05$  &HAe/Be& [2] \\
 AB Aurig\ae & $1.27 \pm 0.025$ & HAe/Be& [5] \\
IM Lup & $1.271 \pm 0.197$ &cTTs& [1] \\
$\mathrm{ESO\ H}\alpha\ 569$ & 1.29 & cTTs  & [3] \\
V4046 Sgr & $1.605 \pm 0.132$& cTTs & [1]\\
UX Ori & $1.79 \pm 0.12$ &HAe/Be& [6] \\
\bottomrule
\end{tabular}
\end{center}
[1]  \citet{ave2018a}, [2] \citet{lag2006a}, [3] \citet{wol2017a} , [4] \citet{rag2012a}, [5] \citet{dif2009a}, [6] \cite{kre2016a} 
\end{table}

The range of $\beta$ used in our simulations is a subset  of these values.
Our  simulations indicate  that the value of $\lambda_{f}$ is very sensitive to the value of the flaring exponent $\beta$, but much less so to the radial  exponent $\alpha$.  For discs that are flatter than the fiducial disc ($\beta < 1.25$),  $\lambda_{f} > 8 \micron$ and for discs with $\beta < 1.20$, $\lambda_{f} > 10 \micron$. That is,  \textit{for flatter discs  the fulcrum point lies beyond the silicate feature}. 

Observations suggest that the surface densities of protoplanetary discs vary approximately as a power law in the inverse cylindrical radius $1/R$ with a positive  exponent $p$.
If the 3D-parametric density equation \ref{eqn:parametric-density} is integrated  over  all $z$  to obtain the surface density  an identical  expression is obtained provided $p = (\alpha - \beta)$.  Integrating the  surface density  over $R$  yields the disc mass  and this additional constraint ensures that  $p < 2$.   
Furthermore,  the density of a protoplanetary disc generally decreases with radius  ($p > 0$),  hence $0 < p <2$.

For the minimum  mass solar  nebula \citet{wei077a} and \citet{hay1981a} both infer $p = 1.5$.  In a series of papers \citet{dal1998a, dal1999a, dal2001a}  showed that for an externally irradiated, passive disc in hydrostatic equilibrium $p \approx  1$.     \citet{hug2008a} fitted millimetre continuum emissions from four nearby disc systems using both the  the truncated power-law and exponentially tapered disc surface density models and found values of $p \approx \gamma  = [ 0.7, 1.3] $ while \citet{and2009a, and2010b}, modelling discs in Ophiuchus using the exponentially tapered model, found $p \approx \gamma = 0.9 \pm 0.2$.

If we fix $\alpha=2.25$,  which is consistent with the  hydrostatic calculations of \citet{dal1999a} and which was used in modelling T Tauri stars with flared discs  by \citet{woo002a, whi003a} and \citet{whi2013a}, then for $\beta=[1.0, 1.3]$, $p=[1.25, 0.95]$ which falls within the range of these values.

The  YSO's SED can be approximately decomposed  into components due to: the star, the inner rim, the dust feature and the outer disc. 
The star and inner rim contribution are largely unchanged by variations in the disk flaring parameter $\beta$. The $10 \mum$ dust feature will be  unaffected in position, but may vary somewhat in magnitude. The component of the SED due to the outer disc however will be affected by the value of the flaring exponent $\beta$. For a flatter disc, the shadow of the puffed inner rim will extend further  radially and the stretched black body produced by the outer disc’s  re-irradiation of the  incoming stellar radiation  will move towards longer wavelengths. For a significantly flared disc, the shadow cast by the inner rim will be truncated closer to the star as the disc rises out of the shadow of the puffed rim more quickly.  Overall, the disk with a greater $\beta$ will intercept more radiation closer to the star and  this will give rise to  a more compressed  outer disc SED component with a peak located at somewhat shorter wavelengths.  If  the inner, puffed rim height is varied this variation in $\beta$ should  cause the SED's of  flatter discs to pivot at longer wavelengths than is the case for more flared discs.
 
In summary, the presence of a fulcrum point in the SED  is {suggestive}  of  a high inclination of the  disc to the line of sight, while the position of the fulcrum point relative to the {$10 \micron$ silicate feature (either in emission or absorption) may be predictive  of how much the disc flares.}

\section{Conclusion}\label{sec:conclusions}
Our radiative transfer simulation studies of \theobj  using Hochunk3D have  established its accretion disc has a fluctuating inner rim height, certainly within the range   $h < H_\mathrm{rim}  < 6 .0 h$.
Furthermore they have shown that:
\begin{enumerate}
	\item The presence of a single fulcrum point $\lambda_{f}$ depends upon the disc being  highly inclined ( $> 70^{\circ}$ )
	\item The inclination cannot be so high that the $10  \micron$ {silicate} feature goes into absorption ($ i < 85$ degrees)
	\item {For our gapped disc model, with a gap between 1 -15 au and with density profile exponents $\alpha = 2.25$ and $\beta=1.25$, we found a single fulcrum point with $\lambda_{f} \approx 8.0 \micron$  near the observed value of $\lambda_{f} =8.5 \micron$}
 	\item Altering the disc accretion rate by itself cannot explain the see-saw variations in flux
	\item The presence or absence of a gap in the disc between 1 and 15 au does not have a strong influence on the position of the fulcrum point - although it does improve the agreement between the simulations and the observations.
	\item The position of the fulcrum point is most strongly influenced by $\beta$ with variations in other parameters not having a strong effect
	\item {Keeping all else fixed, as we reduce the value of $\beta$ below the fiducial value of 1.25, the fulcrum point moves to ever longer wavelengths. For $\beta$ small enough, $\lambda_{f} > 10 \micron$, that is it lies beyond the silicate feature in the SED.}
\end{enumerate}
 

\section*{Acknowledgements}

This work was performed on the gSTAR national facility at Swinburne University of Technology. gSTAR is funded by Swinburne and the Australian Government's Education Investment Fund.\\
{GRB acknowledges the support of a Swinburne University Postgraduate Research Award (SUPRA).}




\bibliographystyle{mnras}
\bibliography{fulcrum} 


%


\bsp	
\label{lastpage}
\end{document}